\documentclass[vecphys]{svmult}
\usepackage{url,graphicx,epsfig}

\def\ledd{L$_{\rm Edd}$}
\def\lx{L$_{\rm X}$}

\def\se{s$^{-1}$ }
\def\porb{P$_{\rm orb}$ }

\def\cm2{cm$^2$ }
\def\ergs{erg~s$^{-1}$ }
\def\gtsima{$\; \buildrel > \over \sim \;$}
\def\ltsima{$\; \buildrel < \over \sim \;$}
\def\prosima{$\; \buildrel \propto \over \sim \;$}
\def\gsim{\lower.5ex\hbox{\gtsima}}
\def\lsim{\lower.5ex\hbox{\ltsima}}
\def\simgt{\lower.5ex\hbox{\gtsima}}
\def\simlt{\lower.5ex\hbox{\ltsima}}
\def\simpr{\lower.5ex\hbox{\prosima}}

\def\ie{{i.e.~}}

\def\etal{{et al.}}

\begin{document}

\title*{Radio emission and jets from microquasars}
\titlerunning{Jets from microquasars}

\author{Elena Gallo}

\authorrunning{Elena Gallo} 

\institute{Hubble Fellow at the MIT Kavli Institute for Astrophysics and Space Research, \\
70 Vassar Street, Bldg 37-685, Cambridge, MA 02139, USA\newline}

\maketitle
\abstract{
To some extent, all Galactic binary systems hosting a compact object are
potential `microquasars',
so much as all galactic nuclei may
have been quasars, once upon a time. The necessary ingredients for a compact object of stellar mass to qualify as a microquasar
seem to be: accretion, rotation and magnetic field. The presence of a black
hole may help, but is not strictly required, since neutron star X-ray binaries
and dwarf novae can be powerful jet sources as well. The
above issues are  
broadly discussed throughout this Chapter, with a a rather trivial question in
mind: {why} do we care?  In other words: are jets a negligible phenomenon in
terms of accretion power, or do they contribute significantly to dissipating 
gravitational potential energy? How do they influence their surroundings? The
latter point is especially relevant in a broader context, as there is
mounting evidence that outflows powered by super-massive black
holes in external galaxies may play a crucial role in regulating the evolution of cosmic 
structures.
Microquasars can also be thought of as a form of quasars for the impatient:
what makes them appealing, despite their low number statistics with respect to
quasars, are the fast variability time-scales. In the first approximation, the
physics of the jet-accretion coupling in the innermost regions should be set
by the mass/size of the accretor: stellar mass objects vary on $10^5-10^8$
times shorter time-scales, making it possible to study variable accretion
modes and related ejection phenomena over average Ph.D. time-scales. At the same time, allowing for a systematic comparison between different classes of
compact objects -- black holes, neutron stars and white dwarfs -- microquasars hold the key to identify and characterize properties that may be unique to, e.g., the
presence (or lack) of an event horizon.}

\section{Radio observations of black holes}
\label{sec:intro}

\begin{figure}[t!]
{\includegraphics[width=1.01\textwidth]{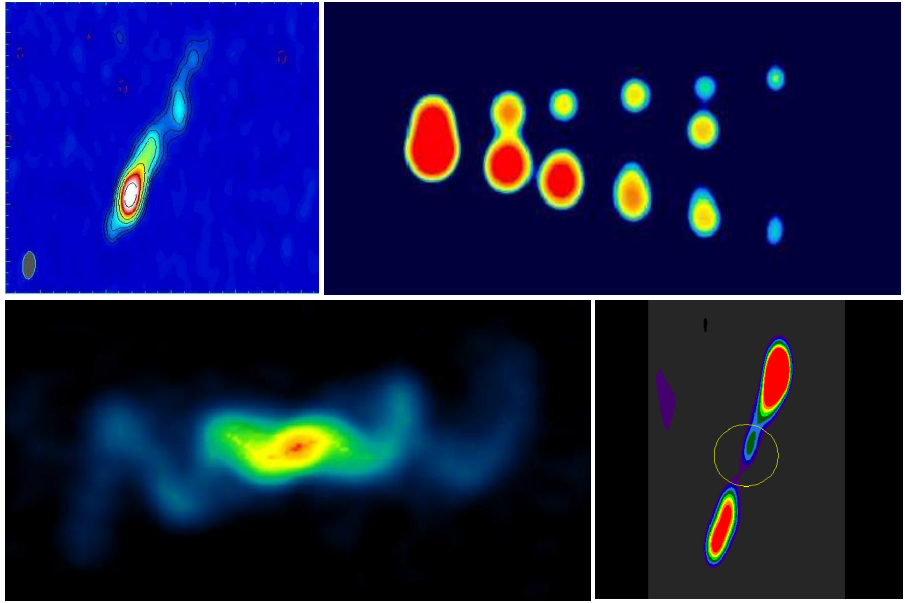}}
\caption{Top left: steady, milliarcsec scale jet from the
high mass black hole X-ray binary Cyg~X-1. From \cite{stirling}. Top right:
transient, arcsec-scale radio jets from the superluminal Galactic jet source
GRS~1915+105. From \cite{mirabel94}. Bottom left: arcsec-scale radio jets from
the first Galactic source discovered: SS~443\index{SS~433}. The binary orbit is almost
edge-on; the precessing accretion disk of SS~433 causes its jets to trace a
`corkscrew' in the sky every 162 days. From \cite{blundell}. Bottom right:
fossil, arcmin-scale radio jets around the Galactic Center black hole in
1E~140.7-2942. From \cite{mirabel92}. }
\label{fig:radiojets}
\end{figure}

The synchrotron nature of the radio emission from X-ray binaries is generally
inferred by the non-thermal spectra and high brightness temperatures. The
latter translate into minimum linear sizes for the radio emitting region which
often exceed the typical orbital separations, making the plasma uncontainable
by any known component of the binary. If coupled to persistent radio flux
levels, this implies the presence of a continuously replenished relativistic
plasma that is flowing out of the system~\cite{hj88,mirrod99,fender06}.
Thanks to aggressive campaigns of multi-wavelength observations of X-ray
binaries in outbursts over the last decade or so, we have now reached a
reasonable understanding of their radio phenomenology in response to global
changes in the accretion mode.  

For the black holes (BHBs), radiatively inefficient, low/hard X-ray states
are associated with flat/slightly inverted radio-to-mm spectra and persistent
radio flux~\cite{fender01} (the reader is referred to Chap. 3 of this book
for a review of X-ray states of black hole X-ray
binaries, as well as:~\cite{mccr,hb05}). In analogy with compact extragalactic radio
sources, the flat spectra are thought to be due to the superimposition of a
number of peaked synchrotron spectra generated along a conical outflow, or
jet, with the emitting plasma becoming progressively thinner at lower
frequencies as it travels away from the jet base~\cite{bk79,kaiser06}. The jet
interpretation has been confirmed by high resolution radio maps of two hard
state BHBs: Cygnus~X-1~\cite{stirling} (Fig.~\ref{fig:radiojets}, top left
panel) and GRS~1915+105~\cite{dhawan,fuchs} are both resolved into elongated
radio sources on milliarcsec-scales (tens of A.U.) implying collimation angles
smaller than a few degrees. Even though no collimated radio jet has been
resolved in any BHB emitting X-rays below a few $\%$ of the Eddington limit,
\ledd, it is widely accepted, by analogy with the two above-mentioned systems,
that the flat radio spectra associated with unresolved radio counterparts of
X-ray binaries are originated in conical outflows. Yet, it remains to be
proven whether such outflow would maintain highly collimated at very low
luminosity levels, in the so called `quiescent' regime (\lx/\ledd$\simlt
10^{-5}$; see Sect.~\ref{ssec:v404}).

Radiatively efficient, high/soft (thermal dominant) X-ray states, on the contrary, are
associated with no {\emph{ flat spectrum}} core radio emission~\cite{fender99};
the core radio fluxes drop by a factor at least 50 with respect to the hard
state 
(e.g. \cite{fender99,corbel04}), which is generally interpreted as the physical
suppression of the jet taking place over this regime. While a number of
sources has been detected in the radio during the soft 
state~\cite{brocksopp02,corbel04,gallo04,brocksopp05}, the common
believe is that these are due to optically thin synchrotron emission -- until
proven otherwise. 

\begin{figure}[t!]
\includegraphics[width=1.1\textwidth,]{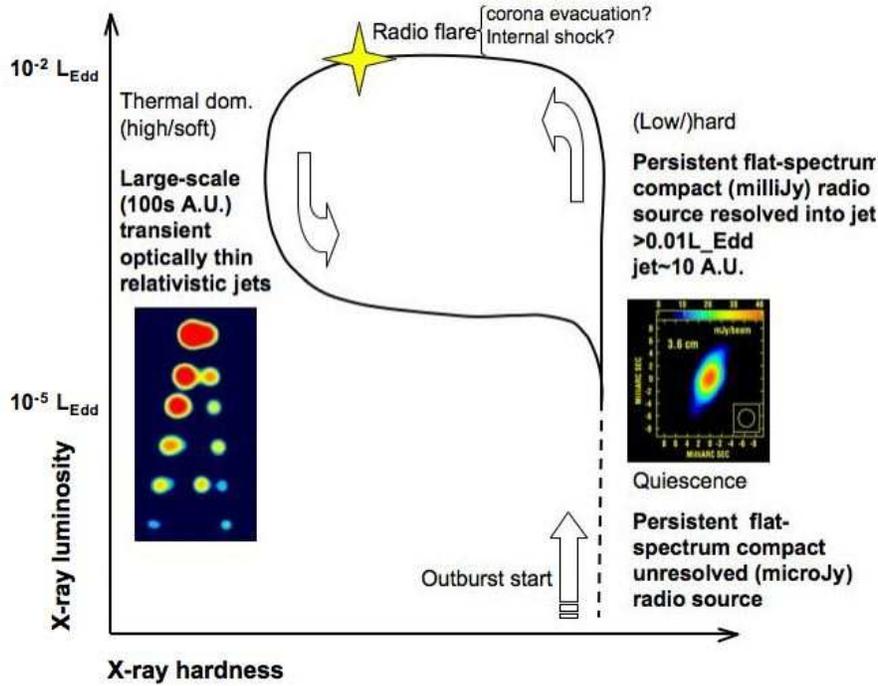}
\caption{Schematic of radio properties of black hole X-ray binaries over
different accretion modes (see Chap. 3 for a description of X-ray
states). See: {\tt{http://www.issibern.ch/teams/proaccretion/Documents.html}}
for a more detailed illustration, and \cite{fbg,fb04} for the relative science
papers. }
\label{fig:turtle} 
\end{figure}

Transient ejections of optically thin radio plasmons moving away from the
binary core in opposite directions are often observed as a result of bright
radio flares associated with hard-to-soft X-ray state transitions
(Fig.~\ref{fig:radiojets}, top right panel). This are surely the most
spectacular kind of jets observed from X-ray binaries, in fact those which
have inspired the term `microquasar'~\cite{mirrod98}. As proven by the case
of GRS~1915+105, and more recently by Cygnus~X-1 as well~\cite{fender06b}, the
same source can produce either kind of jets, persistent/partially
self-absorbed, and transient/optically thin, dependently on the accretion
regime.

\section{Coupling accretion and ejection in black holes}

The question arises weather the flat-spectrum, steady jet and the
optically-thin discrete ejections differ fundamentally or they are different
manifestations of the same phenomenon. This issue has been tentatively
addressed with a phenomenological approach ~\cite{fbg}. Broadly speaking, this
model aims to put together the various pieces of a big puzzle that were
provided to us by years of multi-wavelength monitoring of BHBs, and to do so
under the guiding notion that the jet phenomenon has to be looked at as an
intrinsic part of the accretion process. \cite{fbg} collected as many
information as possible about the very moment when major radio flares occur in
BHBs, and proposed a way to `read them' in connection with the X-ray state
over which they took place as well as the observed jet properties prior and
after the radio flare.  

The study makes use of simultaneous X-ray
(typically Rossi X-ray Timing Explorer, RXTE) and radio (Australia Telescope
Compact Array, ATCA, and/or the Very Large Array, VLA) observations of four
outbursting BHB systems: GRS~1915+105, XTE~J1550-564, GX~339-4 and XTE~J1859+229. X-ray hardness-intensity diagrams (HID) have been constructed for
the various outbursts and linked with the evolution of the jet morphology,
radio luminosity, total power, Lorentz factor and so on.

Figure~\ref{fig:turtle} (naively) illustrates our understanding of the so
called jet-accretion coupling in BH X-ray binaries. It represents the HID of a
well behaved outburst, with the time arrow progressing 
counterclockwise. Starting from the bottom right corner, the system is a
low-luminosity, quiescent X-ray state, producing a (supposedly) mildly
relativistic, persistent outflow, with flat radio spectrum.
Its luminosity starts to increase at all wavelengths, while the X-ray spectrum
remains hard. Around a few $\%$ of the Eddington X-ray luminosity, a sudden
transition is made (top horizontal branch) during which the global properties
of the accretion flow change substantially
(hard-to-soft state transition), while a bright radio flare is observed,
likely due to a sudden ejection episode~\cite{mir98}.  This can be interpreted
\cite{fbg} as the result of the inner radius of a geometrically thin accretion
disk moving inward. The Lorentz factor of the ejected material, due to the
deeper potential well, exceeds that of the hard state jet, causing an internal
shock to propagate through it, and to possibly disrupt it. Once the transition
to the high/soft (thermal dominant) state is made, no core radio emission is
observed, while large scale (hundreds of A.U.), rapidly fading radio plasmons
are often seen moving in opposite direction with respect to the binary system
position, with highly relativistic speed.  Towards the end of the outburst,
the X-ray spectrum starts to become harder, and the compact, flat-spectrum,
core radio source turns on once again. A new cycle begins (with time-scales
that vary greatly from source to source).

The bright radio flare associated with the state transition could coincide
with the very moment in which the hot corona of thermal electrons, responsible
for the X-ray power law in the spectra of hard state BHBs, is accelerated and
ultimately evacuated.  This idea of a sudden evacuation of inner disk material
is not entirely new, and in fact dates back to extensive RXTE observations of
the rapidly varying GRS~1915+105: despite their complexity, the source
spectral changes could be accounted for by the rapid removal of the inner
region of an optically thick accretion disk, followed by a slower
replenishment, with the time-scale for each event set by the extent of the
missing part of the disk~\cite{belloni97a,belloni97b}. Subsequent,
multi-wavelength (radio, infrared and X-ray) monitoring of the same source
suggested a connection between the rapid disappearance and follow up
replenishment of the inner disk seen in the X-rays, with the infrared flare
starting during the recovery from the X-ray dip, when an X-ray spike was
observed~\cite{mir98}.\\

Yet it remains unclear what drives the transition in the radio properties
after the hard X-ray state peak is reached. Specifically, radio observations
of GX~339-4\index{GX~339-4}, XTE~J1550-564\index{XTE~J1550-564} and 
GRS~1915+105\index{GRS~1915+105} indicate that in this phase the
jet spectral index seems to `oscillate' in an odd fashion, from flat to
inverted to optically thin, as if the jet was experiencing some kind of
instability as the X-ray spectrum softens.  Recent, simultaneous RXTE and
INTEGRAL (The INTernational Gamma-Ray Astrophysics Laboratory) observations of
GX~339-4~\cite{belloni06} have shown that the high energy 
cutoff typical of hard state X-ray spectra, either disappears or shifts
towards much higher energies within timescales of hours ($<$8 hr) during
the transition.

Finally, there are at least a couple of recent results that might challenge
some of the premises the unified scheme is based on. The first one is the notion that, for the internal shock scenario
to be at work and give rise to the bright radio flare at the state transition,
whatever is ejected must have a higher velocity with respect to the
pre-existing hard state steady jet. From an observational point of view, this
was supported, one one side, by the lower limits on the transient jets'
Lorentz factors\index{Lorentz factor}, 
typically higher than $\Gamma$=2~\cite{fender03}, and, on the other
hand, by the relative small scatter about the radio/X-ray correlation in hard
state BHBs~\cite{gfp} (see Sect.~\ref{ssec:radx}). The latter has been
challenged on theoretical grounds~\cite{hm04}. 

The second premise has to do with the existence of a geometrically thin
accretion disk in the low/hard state of BHBs. Deep X-ray observations of hard
state BHBs~\cite{jm06a,jm06b,rykoff,tomsick08} (see also \cite{beike}). have shown evidence for a cool
disk component extending close to the innermost stable orbit already during
the bright phases of the hard state, that is prior to the horizontal brunch in
the top panel of Fig.~\ref{fig:turtle} (\lx/\ledd $\simeq 10^{-3}-10^{-2}$).
This challenges the hypothesis of a sudden deepening of the inner disk
potential well as the cause of a high Lorentz factor ejection.  Possibly,
whether the inner disk radius moves close to hole prior or during the
softening of the X-ray spectrum does not play such a crucial role in terms of
jet properties; if so, then the attention should be diverted to a different
component, such as the presence/absence, or the size, of a Comptonizing
corona~\cite{homan01}, which could in fact coincide with the very jet
base~\cite{mnw05}. It is worth mentioning that a recent paper~\cite{liu} gives
theoretical support to the survival of a thin accretion disk down to low
Eddington ratios: within the framework of the disk evaporation model
(e.g. ~\cite{lasota}), it is found that a weak, condensation-fed, {inner} disk
can be present in the hard state of black hole transient systems for
Eddington-scaled luminosities as low as $10^{-3}$ (depending on the magnitude
of the viscosity parameter).  \\

Here I wish to stress that, in addition to solving 
the above-mentioned issues, much work needs to be done
in order to test the consistency of the internal shock scenario as a viable
mechanism to account for the observed changes in the radio properties, given
the observational and theoretical constraints for a given source (such as
emissivities, radio/infrared delays, cooling times, etc.).

Finally, one of the most interesting aspects of the proposed scheme --
assuming that is correct in its general principles -- is obviously its
connection to super-massive BHs in active galactic nuclei (AGN), and
the possibility to mirror different X-ray binary states into different classes
of AGN. This is explored in detail in Chap. 5.

\section{Empirical luminosity correlations\index{empirical correlations}}
\label{sec:corr}

\subsection{Radio/X-ray}
\label{ssec:radx}
In a first attempt to quantify the relative importance of jet vs. disk
emission in BHBs, \cite{gfp} collected quasi-simultaneous radio and X-ray
observations of ten low/hard state sources. This study established the
presence of a tight correlation between the X-ray and the radio luminosity, of
the form L$_{\rm R}\propto$\lx$^{0.7\pm 0.1}$, first quantified for
GX~339-4~\cite{corbel03}. The correlation extends over more than 3 orders of
magnitude in \lx~and breaks down around 2$\%$\ledd, above which the sources
enter the high/soft (thermal dominant) state, and the core radio emission
drops below detectable levels. Given the non-linearity, the ratio
radio-to-X-ray luminosity increases towards quiescence (below a few
$10^{-5}$\ledd). This leads to the hypothesis that the total power output
of quiescent BHBs could be dominated by a radiatively inefficient outflow,
rather than by the local dissipation of gravitational energy in the accretion
flow \cite{fgj,kfm06}.  

Even though strictly simultaneous radio/X-ray observation of the nearest
quiescent BHB, A~0620-00, seems to confirm that the non-linear correlation
holds down to Eddington ratios as low as $10^{-8}$~\cite{gallo06}, many
outliers have been recently been found at higher
luminosities~\cite{corbel04,chaty,brocksopp05,cadolle,rodriguezj,gallo07proc},
casting serious doubts on the universality of this scaling, and the
possibility of relying on the best-fitting relation for estimating other
quantities, such as distance or black hole mass.

\begin{figure}
\centering{
\includegraphics[width=.99\textwidth]{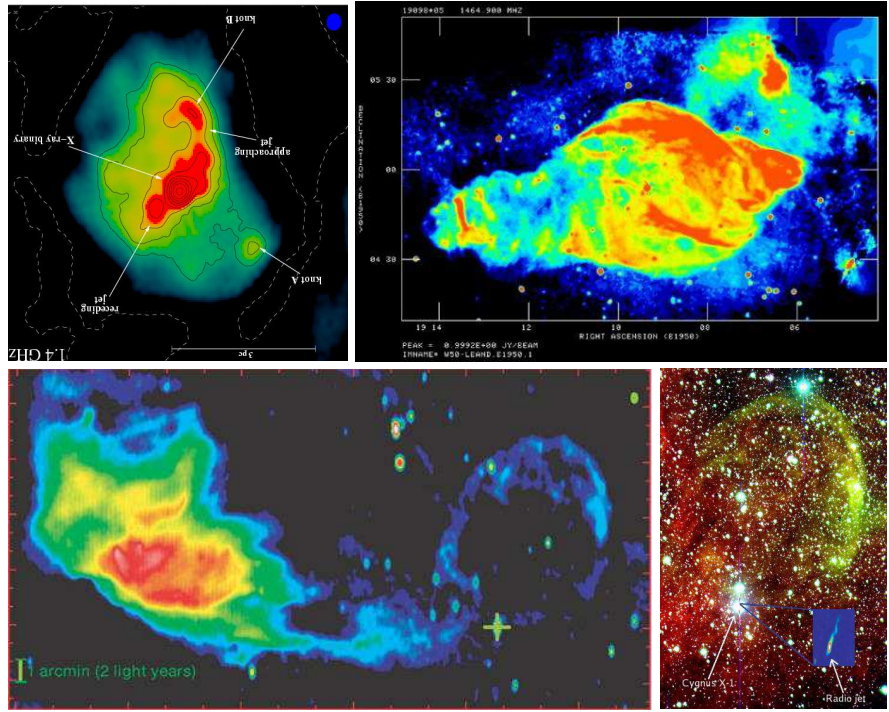}}
\caption{Top left: the jet-powered radio nebula of the `microblazar' Circinus
X-1\index{Cir~X-1}, imaged with the Australia Telescope Compact Array; from
\cite{tudose06}. Top right: W50 nebula surrounding SS~433 (see
Fig.~\ref{fig:radiojets}, bottom left panel, for an image of the
arcsec-scale radio jets). The jets of SS~433 are drilling their way into the
supernova remnant, and give rise to these characteristic
`ears'~\cite{w50}. Bottom left: a jet-powered radio nebula around the black
hole X-ray binary Cygnus~X-1, as seen by the Westerbork Synthesis Radio
Telescope at 1.4 GHz. From 
\cite{gallo05}. Bottom right: the optical counterpart to the Cyg~X-1 nebula,
as observed with the 2.5m Isaac Newton optical telescope. From
\cite{russell07}.}
\label{fig:neb} 
\end{figure}
\subsection{Optical-Infrared/X-ray}
\label{ssec:oir}

The infrared (IR) spectra of BHBs with a low mass donor star are
likely shaped by a number of competing emission mechanisms, most
notably: reprocessing of accretion-powered X-ray and ultraviolet
photons, either by the donor star surface or by the outer accretion
disk, direct thermal emission from the outer disk, and non-thermal
synchrotron emission from a relativistic outflow.
\cite{russell} have collected all the available quasi-simultaneous
optical and near-IR data of a large sample of Galactic X-ray binaries over
different X-ray states. The optical/near-IR (OIR) luminosity of hard/quiescent
BHBs is found to correlate with the X-ray luminosity to the power $\sim$0.6,
consistent with the radio/X-ray correlation slope down to
$10^{-8}$\ledd~\cite{gallo06}.  Combined with the fact that the near-IR
emission is largely suppressed in the soft state, this leads to the conclusion
that, for the BHBs, the spectral break to the optically thin portion of the
jet takes place most likely in the mid-IR (2-40 $\mu$m).
A similar correlation is found in neutron stars (NSs) in the
hard state.
By comparing the observed relations with those expected from models of a
number of emission processes,~\cite{russell} are able to constrain the dominant contribution to the OIR portion of the spectral energy distribution (SED) for the
different classes of X-ray binaries. They conclude that, for hard state BHs at high luminosities (above $10^{-3}$ times the Eddington limit), jets are
contributing 90$\%$ of the near-IR emission. The optical emission could have a substantial jet contribution; however, the optical spectra show a thermal spectrum
indicating X-ray reprocessing in the disk dominates in this regime.
In contrast, X-ray reprocessing from the outer accretion disk dominates the OIR spectra of 
hard state NSs, with possible contributions from the synchrotron emitting jets 
and the viscously heated disk only at very high luminosities.

\begin{figure}
\includegraphics[width=.99\textwidth]{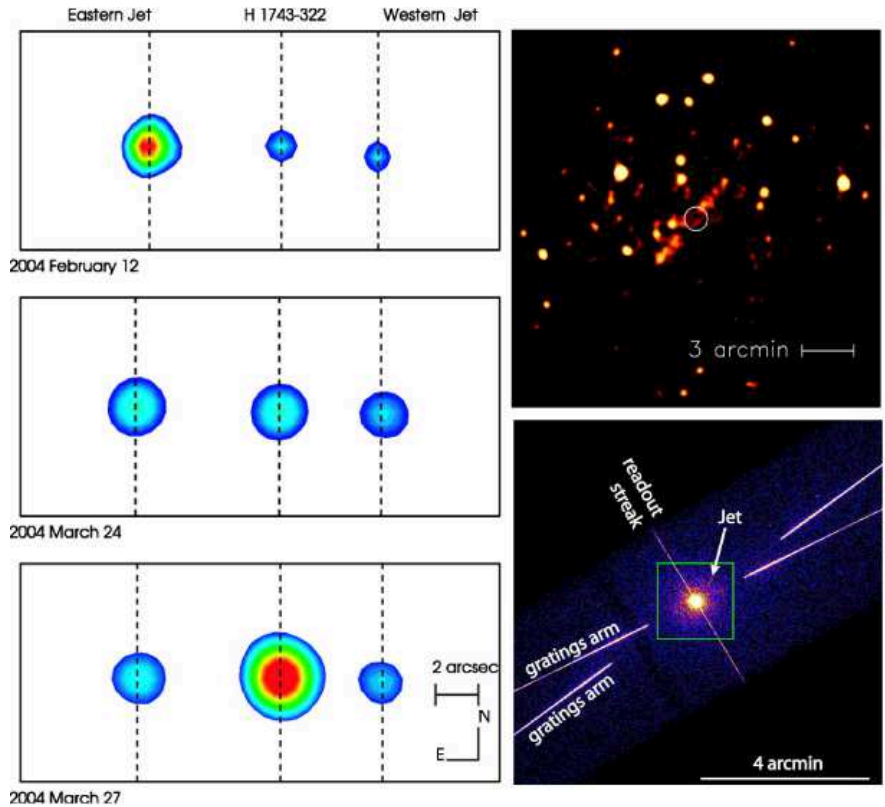}
\caption{Left panels: arcsec-scale, transient X-ray plasmons ejected by the BH
candidate H~1743-322 (Chandra ACIS-S). The detection of optically thin
synchrotron X-ray emission implies in situ particle acceleration up to several
TeV. From \cite{corbel05}. Top right: arcmin-scale fossil X-ray jets in the
field of 4U~1755-33~\cite{angelini} (XMM-Newton image). Bottom right:
arcsec-scale transient X-ray jet from Circinus X-1 seen with the Chandra
gratings. From \cite{heinz07}. The X-ray jet\index{X-ray jets} direction coincide with that of
the ultra-relativistic radio jets \cite{fendercirx1}. }
\label{fig:xjet} 
\end{figure}

\section{Jet-ISM interaction}
\subsection{Jet-driven nebulae}
\label{ssec:nebulae}

It is worth stressing that none of the above scaling relations deals with
actual measurements of the {\em total~}jet power, which is a function of the
observed radio luminosity, corrected for relativistic effects, and of the
unknown radiative efficiency. A fruitful method, borrowed from the AGN
community, is that to constrain the jet power$\times$lifetime product by
looking at its interaction with the surrounding interstellar medium
(ISM). Beside the arcmin-scale, fossil jets around the so called Great
Annihilator (Fig.~\ref{fig:radiojets}, bottom right
panel;~\cite{mirabel92}), a well known case is that of the nebula around the
first Galactic jet source discovered: SS~433\index{SS~433}. The `ears' of W50
(Fig.~\ref{fig:neb}, top right panel) act as an effective calorimeter for
the jets' mechanical power, which is estimated to be greater than $10^{39}$
\ergs~\cite{begelman80}.  Similarly, the neutron star X-ray binary in Circinus
X-1 is embedded in an extended, jet-driven radio nebula\index{radio nebulae} (see
Fig.~\ref{fig:neb}, top left panel). In this particular case, it is likely
that we are actually looking toward the central X-ray binary system through
the jet-powered radio lobe, making this the only known case of a Galactic {\emph{
microblazar}}. Results from modeling suggest an age for the nebula of $\simlt
10^5$ yr and a corresponding time-averaged jet power in excess of $10^{35}$
\ergs. During flaring episodes, the instantaneous jet power may reach values
of similar magnitude to the X-ray luminosity~\cite{tudose06}.

More recently, a low surface brightness arc of radio/optical emission has been
discovered around Cygnus~X-1~\cite{gallo05,russell07}
(Fig.~\ref{fig:neb}, bottom panels) and interpreted in terms of a shocked compressed
hollow sphere of free-free emitting gas driven by an under-luminous
synchrotron lobe inflated by the jet of Cygnus~X-1.  The lack of a visible
counter arc is ascribed to the lower interstellar matter density in the
opposite direction. In fact, there exist to date relatively few cases where
jet-ISM interactions have been directly observed \cite{james08}. Carrying the
analogy of AGN jet-ISM interactions over to microquasars, it has been argued
that microquasars are located, dynamically speaking, in much more tenuous
atmospheres. As a consequence, compared to AGN, microquasar jets require
particularly dense environments in order to produce visible signs of
interaction with the surroundings \cite{heinz06,heinz03}.

\subsection{X-ray jets\index{X-ray jets}}
\label{ssec:xjet}

It is well known that, in AGN, optical and X-ray jets are also frequently
seen. With the exception of the large-scale (tens of pc) diffuse X-ray
emission detected from the X-ray binary SS~433 with the Einstein
Observatory~\cite{seward80}, X-ray jets were not seen for Galactic systems
prior to the launch of the Chandra X-Ray Observatory in 1999. With a large
improvement in angular resolution over previous missions, Chandra detected
arcsec-scale ($\sim$0.025 pc) X-ray jets for the first time in
SS~433~\cite{marshall02}. This was the first of a number of discoveries.

Perhaps the most extreme case in terms of energetics has been the detection of
decelerating arcsec-scale X-ray (and radio) jets in the microquasar XTE~J1550-564, a few years after the ejection
event~\cite{corbel02,tomsick03jet,kaaret03}. The detection of optically thin
synchrotron X-ray emission from discrete ejection events implies in situ
particle acceleration up to several TeV, possibly due to interaction of the
jets with the interstellar medium.  More recently, a similar large scale jet
has been reported in H~1743-322\index{H~1743-322} \cite{corbel05} (Fig.~\ref{fig:xjet}, left 
panels).  As for XTE~J1550-564\index{XTE~J1550-564}, the spectral 
energy distribution of the jets
during the decay phase is consistent with a classical synchrotron spectrum of
a single electron distribution from radio up to X-rays, implying the
production of very high energy ($>$10 TeV) particles in those jets.

Another interesting example is that of a fossil arcmin-scale X-ray jet seen by
the{ XMM-Newton} telescope in the surroundings of 4U~1755-33~\cite{angelini}\index{4U~1755-33}
(Fig.~\ref{fig:xjet}, top right panel).  Finally, evidence for a transient
X-ray jet has been recently claimed in the neutron star Circinus~X-1\index{Cir~X-1} during a
50 ks Chandra gratings observation (Fig.~\ref{fig:xjet}, bottom right 
panel), taken during a low flux
state~\cite{heinz07}.  The direction of this X-ray feature is consistent with
the direction of the northwestern jet seen in the radio \cite{fendercirx1},
suggesting that it originates either in the jet itself or in the shock that
the jet is driving into its environment. The inferred jet kinetic power is
significantly larger than the minimum power required for the jet to inflate
the large-scale radio nebula.

\section{Quiescence (to eject, or not to eject?)\index{quiescent emission}}
\label{sec:qui}
The role of outflows is especially interesting at very low X-ray luminosities,
in the so called `quiescent' regime, i.e. below a few
$10^{-5}$\ledd. Persistent, steady radio counterparts to BHBs appear to
survive down to low quiescent X-ray luminosities~(as low as $10^{-8.5}$
\ledd~\cite{gallo06}), even though sensitivity limitations on current radio
telescopes make it extremely difficult to reach the signal-to-noise ratios
required to assess their presence for systems further than 2 kpc or so.

In the context of X-ray binaries, as well as super-massive black
holes, the term `jet' is typically used as a synonymous for
relativistic outflow of plasma and implies a high degree of
collimation. As a matter of fact, high spatial resolution radio
observations of BHBs in the hard state have resolved 
highly collimated structures in two systems only: Cyg
X-1~\cite{stirling} and GRS~1915+105~\cite{dhawan,fuchs} are both resolved
into elongated radio sources on milliarcsec scales -- that is tens of
A.U. -- implying collimation angles smaller than a few degrees on
much larger scales than the orbital separation.  Both systems display a
relatively high X-ray (and radio) luminosity, with GRS~1915+105 being
persistently close to the Eddington luminosity~\cite{fb04}, and Cyg
X-1 displaying a bolometric X-ray luminosity around 2$\%$\ledd~\cite{disalvo01}.

This, however, should not be taken as evidence against collimated jets in 
low luminosity, quiescent systems: because of sensitivity limitations on current high
resolution radio arrays, resolving a radio jet at microJy level 
simply constitutes an observational challenge. In addition, at such low
levels, the radio flux could be easily contaminated by synchrotron emission
from the donor star.

In principle, the presence of a collimated outflow can also be inferred by its long-term
action on the local interstellar medium, as in the case of the hard state BHBs
1E~1740.7-2942\index{1E~1740.7-2942}
and GRS~1758-258\index{GRS~1758-258}, both associated with arcmin-scale radio
lobes~\cite{mirabel92,marti98}.  Further indications can come from the stability
in the orientation of the electric vector in the radio polarization maps, as
observed in the case of
GX~339$-$4 over a two year period~\cite{corbel00}. This constant position
angle, being the same as the sky position angle of the large-scale, optically
thin radio jet powered by GX~339-4 after its 2002 outburst~\cite{gallo04},
clearly indicates a favoured ejection axis in the system. However, all three
systems emit X-rays at `intermediate' luminosities ($10^{-3}$-$10^{-2}$\ledd),
and tell us little about outflows from quiescent BHs.  On the other hand,
failure to image a collimated structure in the hard state of
XTE~J1118+480\index{XTE~J1118+480} down to a synthesized beam of 0.6$\times$1.0 mas$^{2}$ at 8.4
GHz~\cite{mirabel01} poses a challenge to the {collimated} jet interpretation, even
though XTE~J1118+408 was observed at roughly one order of magnitude lower
luminosity with respect to e.g. Cyg~X-1 ($10^{-3}$ \ledd). Under the (naive)
assumption that the jet size scales as the
radiated power, one could expect the jet of XTE~J1118+408 to be roughly ten
times smaller than that of Cyg~X-1 (which is 2$\times$6 mas$^{2}$ at 9 GHz, at
about the same distance), \ie still point-like at Very Long Based Array (VLBA)
scales~\cite{mirabel01}.

In fact, \cite{garcia03} have pointed out that long period ($\simgt 1$ day)
BHBs undergoing outbursts tend to be associated with spatially resolved
optically thin radio ejections, while short period systems would be associated
with unresolved, and hence physically smaller, radio ejections. If a common
production mechanism is at work in optically thick and optically thin BHB
jets, then the above arguments should apply to steady optically
thick jets as well, providing an alternative explanation to the unresolved
radio emission of XTE~J1118+480 (which, with its 4 hr orbital period, is one
of the shortest
known).  By analogy, a bright, long period system, like for instance V404~Cyg, might be expected to have a more extended optically thick jet. This is
further explored in the next Section.

\subsection{The brightest: V404 Cyg\index{V404 Cyg}}
\label{ssec:v404}

\begin{figure}
\includegraphics[width=.4\textwidth]{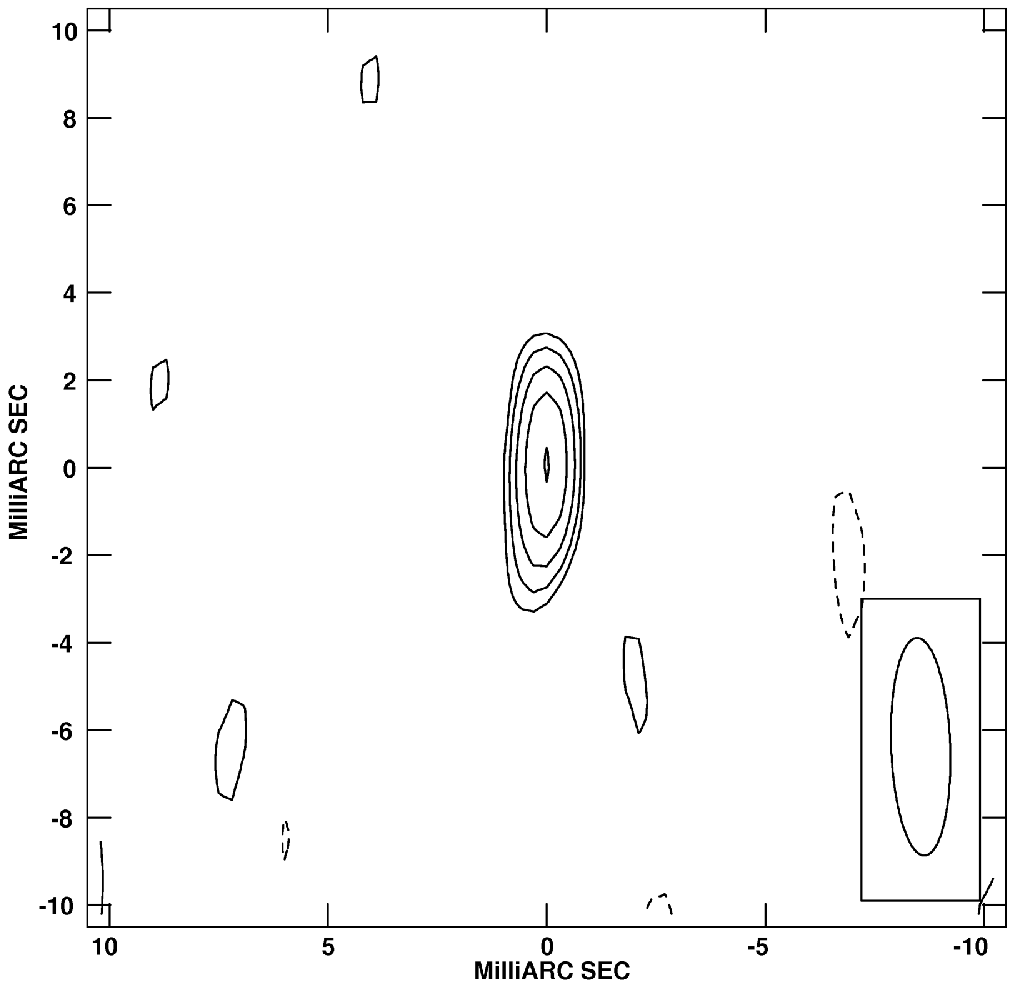}
\includegraphics[width=.6\textwidth]{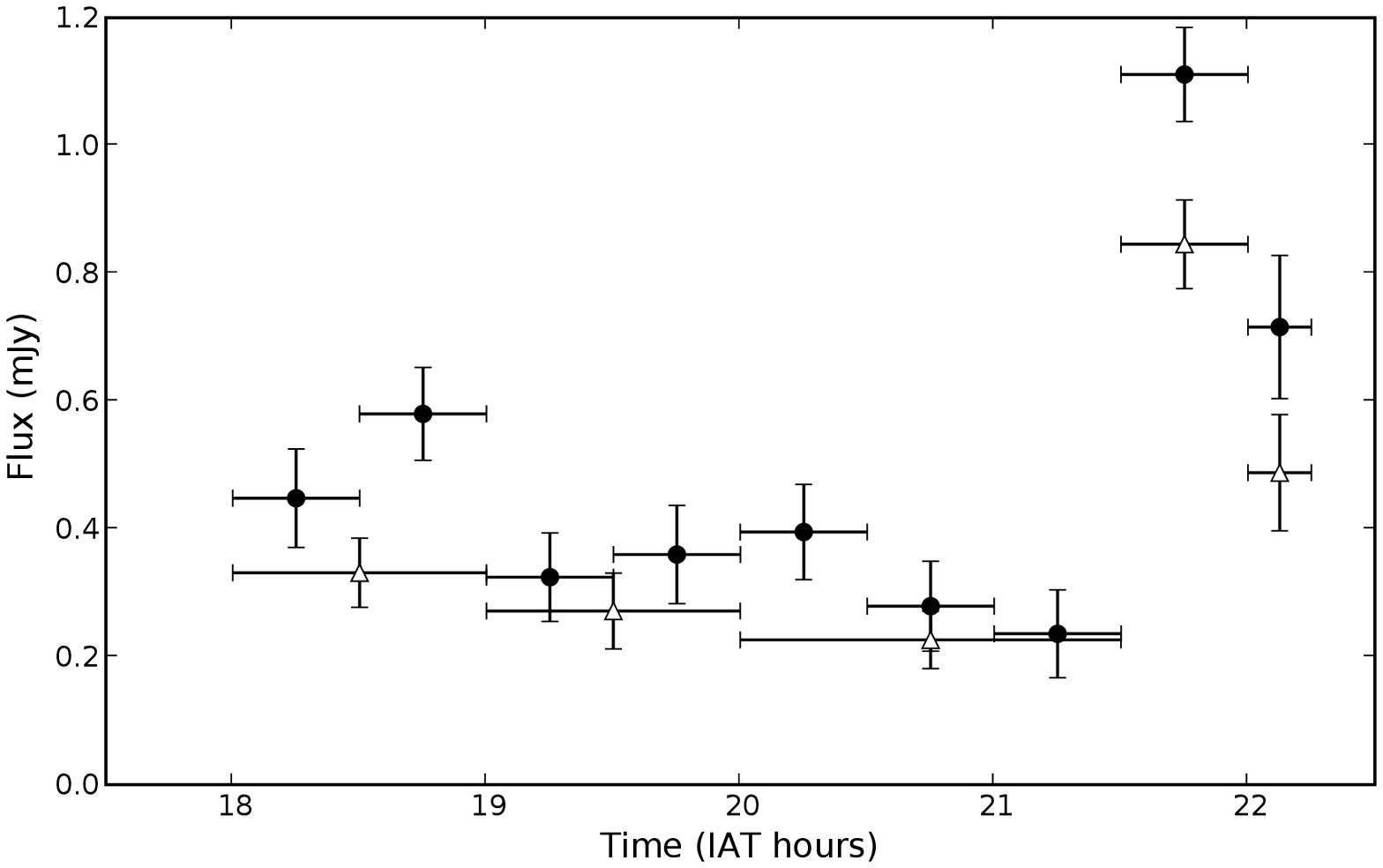}
\caption{Left: the quiescent black hole V404 Cyg was observed  with the High
Sensitivity Array for 4.25 hr at 8.4 GHz. The radio source is unresolved,
yielding an upper limit of 5.2 A.U. to the size of the emitting region. Right:
the source radio light curve during the observation, showing a flare with 30
min rising time. From \cite{jmj08}. }
\label{fig:hsa} 
\end{figure}

In order to eventually resolve the radio counterpart to a quiescent X-ray
binary, the black hole V404~Cyg was observed with the High Sensitivity Array
(HSA, composed of: VLBA plus the Green Bank Telescope, Effelsberg and the
phased VLA) in 2007 December, for 4.25 hr~\cite{jmj08}. These observations
failed to resolve the radio source (Fig.~\ref{fig:hsa}, left panel),
yielding an upper limit of 1.3 milliarcsec on the 8.4 GHz source size -- 5.2\,A.U. at
4 kpc -- and a corresponding lower limit of $7\times10^6$\,K on its brightness
temperature (confirming that the radio emission must be
non-thermal). Interestingly, the inferred upper limit on the radio source size
is already 2 times smaller than the steady jet resolved in Cyg~X-1 with the
VLBA (Fig.~\ref{fig:radiojets}, top left panel). 

A small flare was detected from V404~Cyg 
with both the VLA and the VLBA, on a timescale of 1 hr (Fig.~\ref{fig:hsa},
right panel), in which the source flux density rose by a factor of 3.
As the brightest black hole X-ray binary in quiescence (few $10^{-5}$ \ledd~in
X-rays, 0.4 mJy at GHz frequencies), V404~Cyg is the only source in which such
flaring activity has been detected in the quiescent state (the 
flare is certainly intrinsic to the source, and can not be caused by
interstellar scintillation~\cite{jmj08}). The question of
whether flares are unique to this source, or are common in such
systems, has direct implications for the nature of the accretion process at low
luminosities.  Observations of XTE~J1118+480 and GX~339-4 in their hard states have shown evidence for fast
variability in the optical and X-ray bands~\cite{hynes00,hynes03,hynes04},
with properties inconsistent with X-ray reprocessing and more indicative of
synchrotron variability. The timescales are shorter than seen in V404~Cyg, and
there were no high-time resolution radio data for comparison.  A better
comparison is the quiescent BH in the center of the milky Way, Sgr~A*. It
shows radio and infrared flaring activity, which has been explained as
adiabatic expansion of a self-absorbed transient population of relativistic
electrons~\cite{yz06}.  Whether there is a truly steady
underlying jet as assumed by standard jet models \cite{bk79,kaiser06} or whether the
emission is composed of multiple overlapping flares \cite{kaisersun} remains
to be determined. More sensitive instruments are necessary to probe these
short-timescale ares and determine the nature of the quiescent jet emission
(see Sect.~\ref{sec:future}).

\subsection{The faintest: A~0620-00\index{A~0620-00}}
\label{ssec:a0620}
In spite of the large degree of uncertainty on the overall geometry of the
accretion flow in this regime, there is general agreement that the X-ray
emission in quiescent BHBs comes from high-energy electrons near the BH.  The
SEDs of quiescent BHBs, as well as low-luminosity AGN are often examined in
the context of the advection-dominated accretion flow (ADAF)
solution~\cite{ny94,ny95}, whereby the low X-ray luminosities are due to
a highly reduced radiative efficiency, and most of the liberated accretion
power disappears into the horizon. Here, due to the low densities, a two
temperature inflow develops, a significant fraction of the viscously
dissipated energy remains locked up in the ions as heat, and is advected
inward, effectively adding to the BH mass. The ADAF model successfully
accounts for the overall shape of the UV-optical-X-ray spectra of quiescent
BHBs (see e.g. \cite{mcc03} for an application to the high quality data of XTE~J1118+480).  Nevertheless, alternative suggestions are worth being considered.
\cite{bb99} elaborated an `adiabatic inflow-outflow solution' (ADIOS), in
which the excess energy and angular momentum is lost to an outflow at all
radii; the final accretion rate into the hole may be only a tiny fraction of
the mass supply at large radii.

\begin{figure}
\hspace{0.5cm}\includegraphics[width=0.8\textwidth]{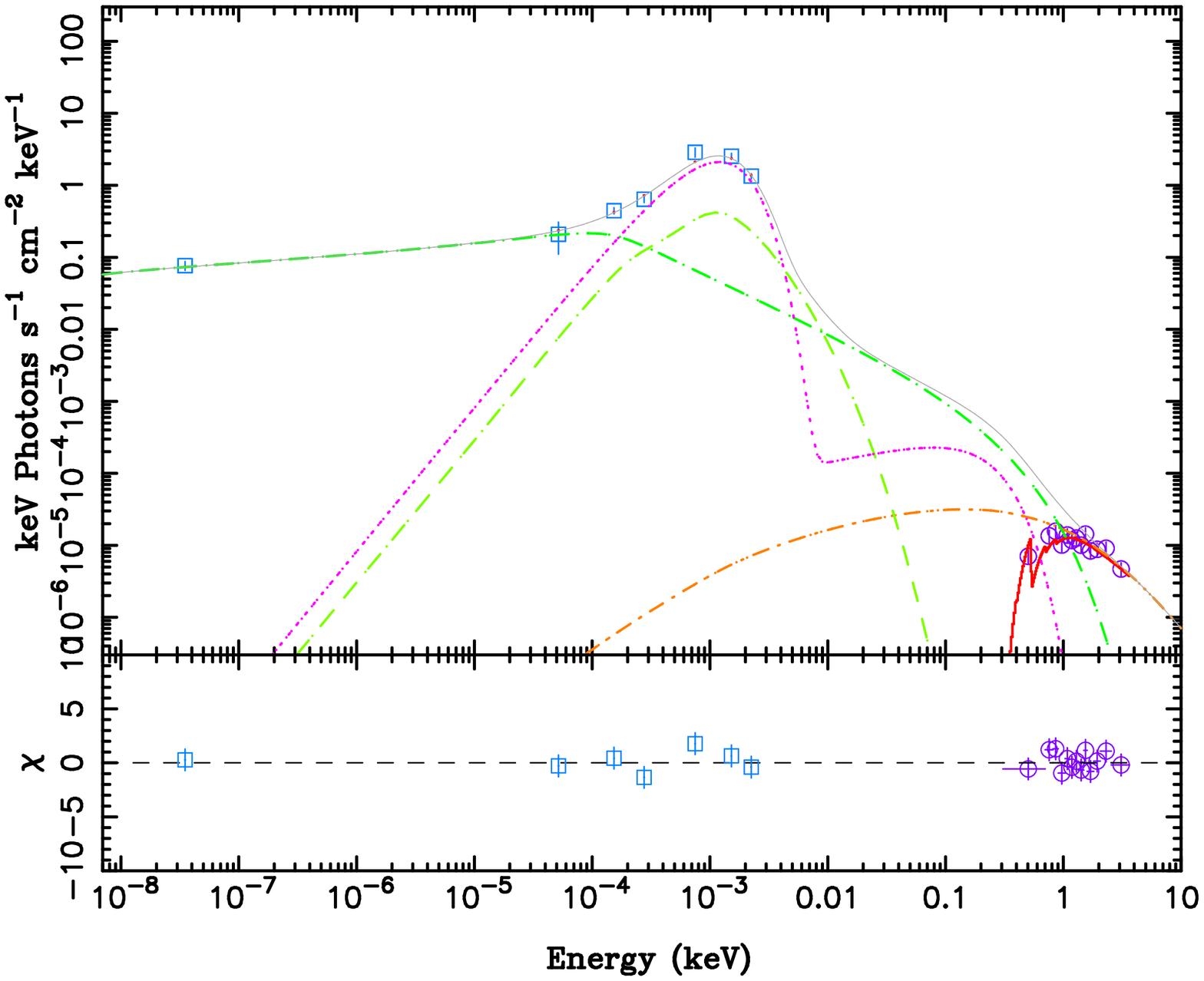}
\hspace{2.5cm}\includegraphics[width=1.5\textwidth]{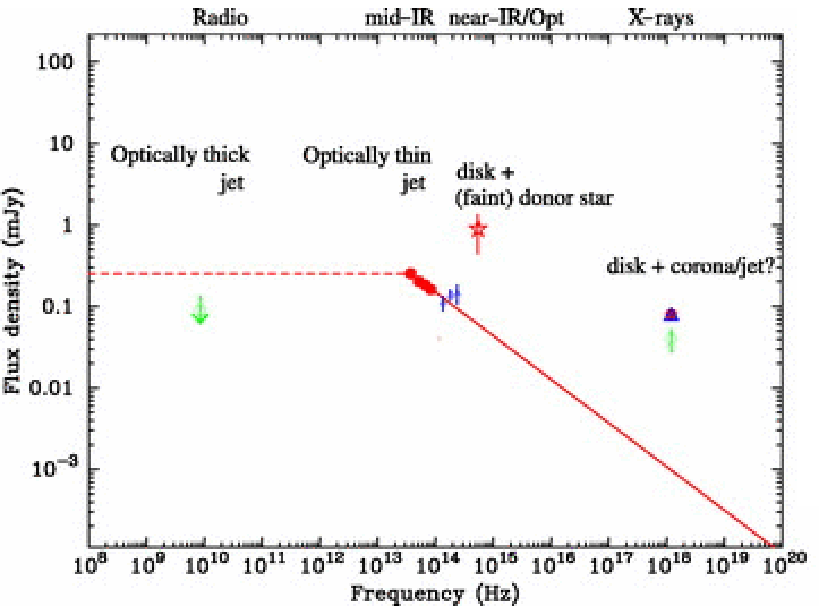}
\caption{Top: radio-to-X-ray SED of the quiescent black hole X-ray binary A~0620-00, fit with a `maximally-dominated' jet model.
Solid gray line: total spectrum; dot-long-dashed yellow line: pre-acceleration inner jet synchrotron emission; dotted green line: post-acceleration outer jet synchrotron; triple dot-dashed orange line: Compton emission from the inner jet, including external disk photons as well as synchrotron self-Compton; double-dot-dashed magenta line: thermal multicolor-blackbody disk model plus single blackbody representing the star. The symbols represent the data, while the solid red line is the model fit in detector space. From~\cite{gallo07}. Bottom: evidence for optically thin synchrotron emission from a jet in the NS 4U~0614+091. By comparison, if the mid-IR excess detected in the BHs A~0620-00 is indeed due to jet emission, this means that the power content of the BH jet is at least 10 times higher than in the NS, where the jet spectrum breaks to optically thin already at mid-IR frequencies. From~\cite{migliari06}.}
\label{fig:spi}
\end{figure}

Alternatively, building on the work by~\cite{fb95} on AGN jets, a jet model
has been proposed for hard state BHBs
\cite{mff01,mff03,mnw05}. Figure~\ref{fig:spi}, top panel, shows a fit to the
radio-to-X-ray SED of A~0620-00, the lowest Eddington-ratio BHB with a
detected radio counterpart (\lx/\ledd$\simeq 10^{-8}$; F$_{\rm 8~GHz}$=50
$\mu$Jy) with such a `maximally jet-dominated' model~\cite{gallo07}.  This is
the first time that such a complex model was applied in the context of
quiescent BHBs, and with the strong constraints on the jet break frequency
cut-off provided by the Spitzer Space Telescope data in the mid-IR regime (see
Sect.~\ref{sec:midir} for more details).
In terms of best-fitting parameters, the major difference with respect to
higher luminosity sources for which this model has been tested
\cite{mff01,mff03} (see Chap. 6) is in the value of the
acceleration parameter $f$ compared to the local cooling rates, which turns
out to be two orders of magnitude lower for A~0620-00.  This `weak
acceleration' scenario is reminiscent of the Galactic Center super-massive BH
Sgr~A*\index{Sgr~A*}. Within this framework, the SED of Sgr~A* does not require a power law
of optically thin synchrotron emission after the break from its flat/inverted
radio spectrum. Therefore, if the radiating particles have a power-law
distribution, it must be so steep as to be indistinguishable from a Maxwellian
in the optically thin regime. In this respect, they must be only weakly
accelerated.  In the framework of a jet-dominated model for the quiescent
regime, it appears that something similar, albeit less extreme,
is occurring in the quiescent BHB A~0620-00; either scenario implies that
acceleration in the jets is inefficient at $10^{-9}-10^{-8}$ \ledd.

\section{Neutron stars\index{neutron-star jets}}
\label{sec:ns}
The mechanism(s) of jet production, from an {\emph{observational}} point of view,
remains essentially unconstrained. While in the case of super-massive BHs in
AGN it is often implicitly assumed that the jets extract their energy from the
rotation of the centrally spinning black hole via large scale magnetic field
lines that thread the horizon, in the case of X-ray binaries, the relatively
low (lower limit on the) jets' Lorentz factors \cite{fender03} do not appear to require
especially efficient launching mechanisms. While on the `experimental' side,
substantial improvements are being made with fully relativistic
magneto-hydrodynamic simulations (e.g. \cite{mck06}, and references therein), from the
observer perspective it seems that a fruitful -- and yet
relatively unexplored -- path to pursue is that to compare in a systematic
fashion the properties of jets in black hole systems to that of e.g. low
magnetic field NSs. 

\begin{figure}
\includegraphics[width=.95\textwidth]{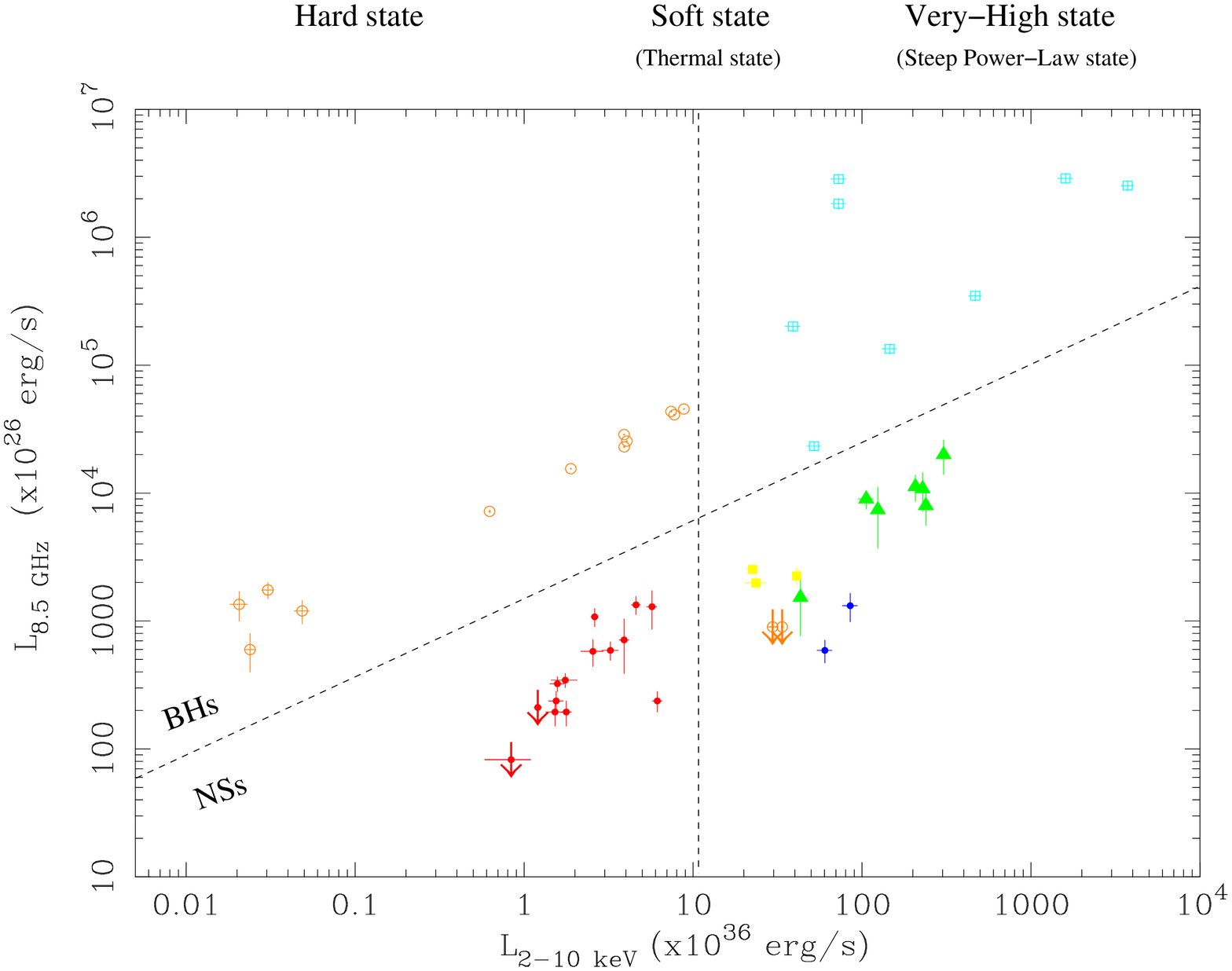}
\includegraphics[width=.45\textwidth]{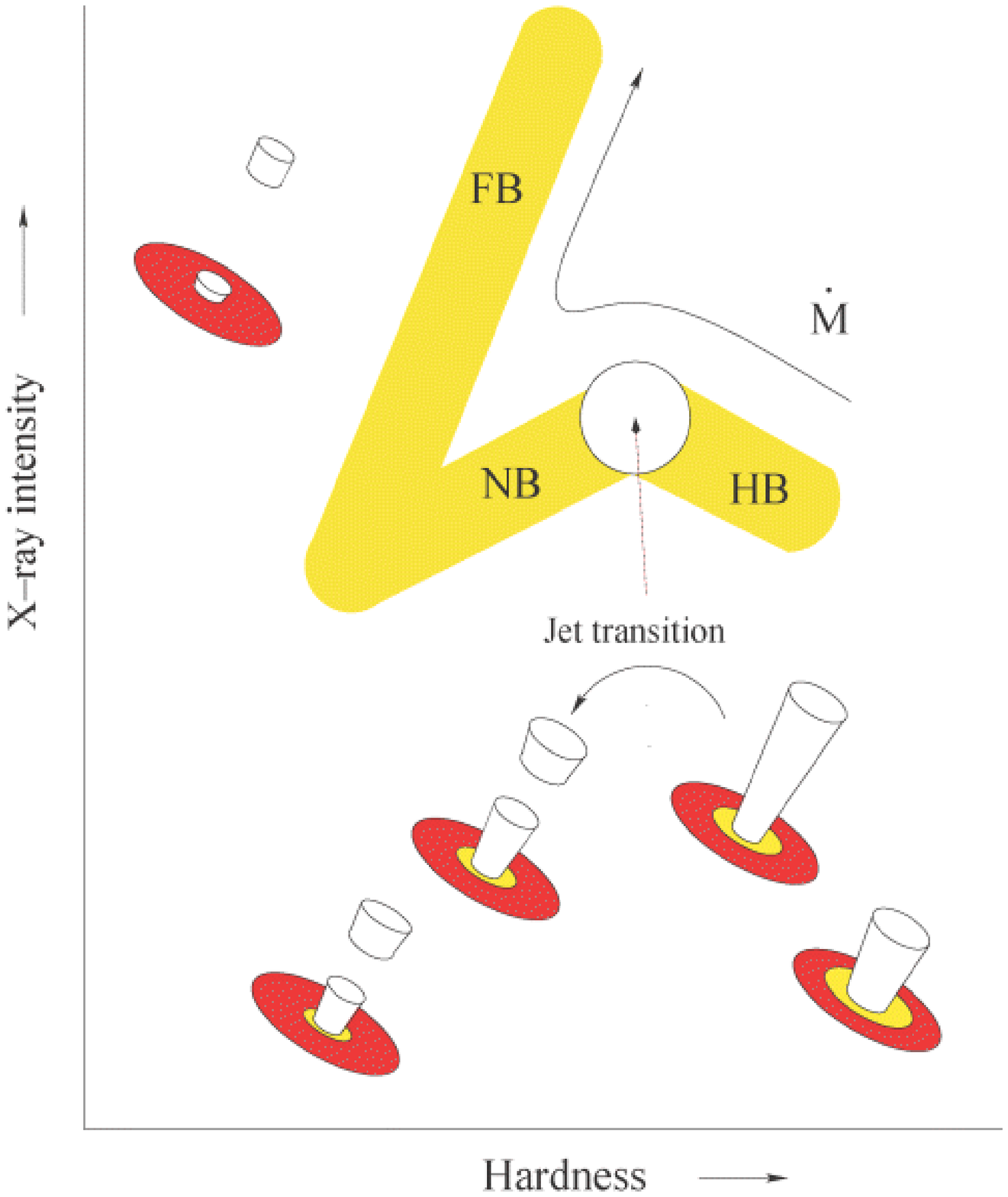}~~~\includegraphics[width=.45\textwidth]{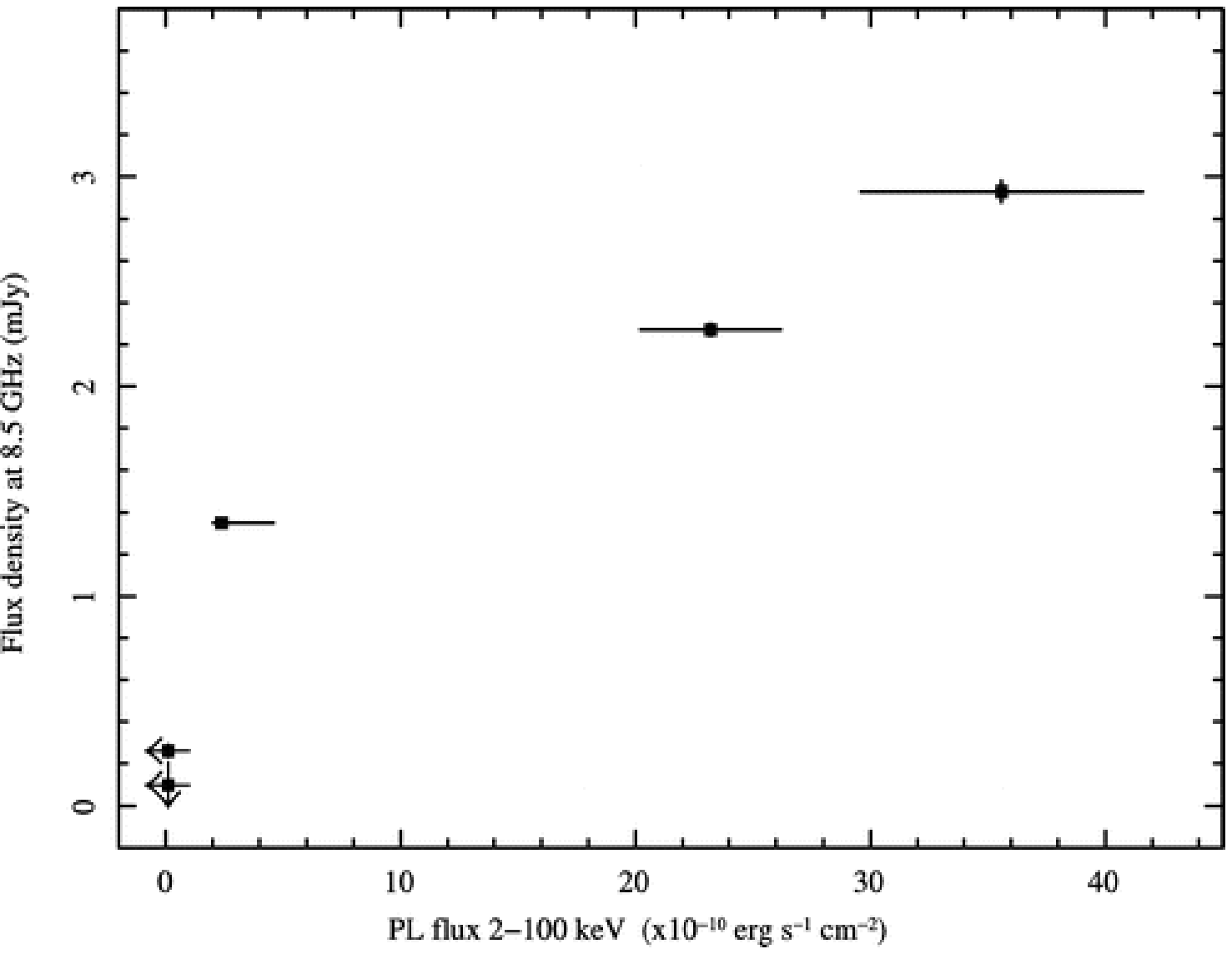}
\caption{Top: radio vs. X-ray luminosity for (2 representative) \index{empirical correlations}
black holes (top left) and neutron stars (bottom right). On average, neutron
stars are 30 times fainter in the radio band. From \cite{mf06}. 
Bottom left: sketch of the disk-jet coupling in Z-type sources, which
are believed to be the NS equivalents of the BH GRS~1915+105, constantly
emitting near \ledd~and producing powerful jets
associated with rapid state transitions. From \cite{mf06}. 
Bottom right: a correlation
between the core radio flux density and the X-ray flux in the hard X-ray in
the Z source GX~17+2. From \cite{migliari07}. }
\label{fig:ns}
\end{figure}

A comprehensive study comparing the radio properties of BHs and
NSs~\cite{mf06} has highlighted a number of relevant difference/similarities
(see Fig.~\ref{fig:ns}): {\it i)} Below a few per cent of the Eddington
luminosity (in the hard, radiatively inefficient states) both BHs and NSs
produce steady compact jets, while transient jets are associated with variable
sources/flaring activity at the highest luminosities.~~{\it ii)} For a given
X-ray luminosity, the NSs are less radio loud, typically by a factor of 30
(Fig.~\ref{fig:ns}, top panel).~~{\it iii)} Unlike BHs, NSs do not show a
strong suppression of radio emission in the soft/thermal dominant state.~~{\it
iv)} Hard state NSs seem to exhibit a much steeper correlation between radio
and X-ray luminosities.\\

Highly accreting NSs, called Z-type (Fig.~\ref{fig:ns}, bottom left panel),
show periodic X-ray state transitions on timescales of a few days. Z sources
can be considered the NS counterparts of transient, strongly accreting BHs
such as GRS~1915+105. At the same time, they are known to display hard
non-thermal tails in the X-ray energy spectra. The physical origin of this
non-thermal component is still an area of controversy, with two main competing
models: 1. inverse Compton scattering from a non-thermal electron population
in a corona~\cite{poutanencoppi}, and 2. bulk motion Comptonization~\cite{titbulk}.  Recently, simultaneous
VLA/RXTE observations of the Z source GX~17+2\index{GX~17+2} have shown a positive
correlation between the radio emission and the hard tail power-law X-ray flux
in this system (Fig.~\ref{fig:ns}, bottom right panel,
\cite{migliari07}). If further confirmed with a larger sample and improved
statistics, this relation would point to a common mechanism for the production
of the jet and hard X-ray tails.

\section{Jet power: the mid-IR leverage}
\label{sec:midir}

\index{jet power}Observations of hard state BHBs
have established that synchrotron emission from the steady jets extends
all the way from the radio to the mm-band \cite{fender01j}, above which the break to the
optically thin portion of the spectrum is thought to occur. 
However, even for the highest quality SED, disentangling the relative contributions of inflow vs. outflow to the
radiation spectrum and global accretion energy budget can be quite
challenging, as exemplified by the emblematic case of
XTE~J1118+480~\cite{mcc03,mff03,yuan05}\index{XTE~J1118+480}.  
Estimates of the total jet power
based on its radiation spectrum depend crucially on the assumed frequency at
which the flat, partially self-absorbed spectrum turns and becomes optically
thin, as the jet `radiative efficiency' depends ultimately on the location of
the high-energy cutoff induced by the higher synchrotron cooling rate of the
most energetic particles.
From a theoretical point of view, the `break frequency', here defined as the
frequency at which the partially self-absorbed jet becomes optically thin, is
inversely proportional to the BH mass: as jet spectral breaks are often
observed in the GHz/sub-mm regime in AGN, they are expected to occur in the
IR-optical band for $10^{5-7}$ times lighter objects.  We know however from
observations of {GX~339-4}, the only BHB where the optically thin jet
spectrum has been perhaps observed
\cite{cf02,homan05}, that the exact break frequency can 
vary with the overall luminosity, possibly reflecting changes in the
magnetic field energy density, particle density and mass loading at
the jet base. Determining the location of
the jet break as a function of the bolometric luminosity is also important
to assess the synchrotron contribution to the hard X-ray band.  
As an example, that the optically thin jet IR-emission in GX~339-4
connects smoothly with the hard X-ray power law has led to challenge
the `standard' Comptonization scenario for the hard X-ray state~\cite{mff03}.\\

Because of the low flux levels expected from the jets in this regime (10--100s
of $\mu$Jy, based on extrapolation from the radio band), combined with the
companion star/outer disk contamination at near-IR frequencies, the
sensitivity and leverage offered by Spitzer is crucial in
order to determine the location of the jet break.  In fact, Spitzer
observations of three quiescent black hole X-ray binaries, with the Multi-band
Imaging Photometer (MIPS), have shown evidence for excess emission with respect
to the Rayleigh-Jeans tail of the companion star between 8--24 $\mu$m. This
excess, which has been interpreted as due to thermal emission from cool
circumbinary material~\cite{mm06}, is also consistent with the extrapolation
of the measured radio flux assuming a slightly inverted spectrum, typical of
partially self-absorbed synchrotron emission from a conical
jet~\cite{gallo07}.  If so, then the jet synchrotron luminosity exceeds the
measured X-ray luminosity by a factor of a few in these systems. Accordingly,
the mechanical power stored in the jet exceeds the bolometric X-ray luminosity
at least by 4 orders of magnitude (based on kinetic luminosity function of
Galactic X-ray binary jets~\cite{heinzgrimm}).\\

Despite their relative faintness with respect to BHs at low (radio)
frequencies, the same multi-wavelength approach can be undertaken in NSs, in
the hope to detect excess mid-IR emission.  Indeed, coordinated radio (VLA),
mid-IT (Spitzer), optical (YALO SMARTS) and X-ray (RXTE) observations, have
yielded the first spectroscopical evidence for the presence of a steady jet in
a low-luminosity, ultra-compact\footnote{That is, with orbital period shorter than 1 hour.} {neutron star} X-ray binary (4U~0614+091
\cite{migliari06}).  The Spitzer data (Fig.~\ref{fig:spi}, bottom panel) show
a neat optically {\emph{thin}} synchrotron spectrum ($F_{\nu}\propto
\nu^{-0.6}$), indicating that the jet break occurs at much lower
frequencies in this neutron star system with respect to black hole X-ray
binaries. As a consequence, unlike in the black holes, the jet can not
possibly contribute significantly to the X-ray emission.

\begin{figure}
\centering{\includegraphics[width=.95\textwidth]{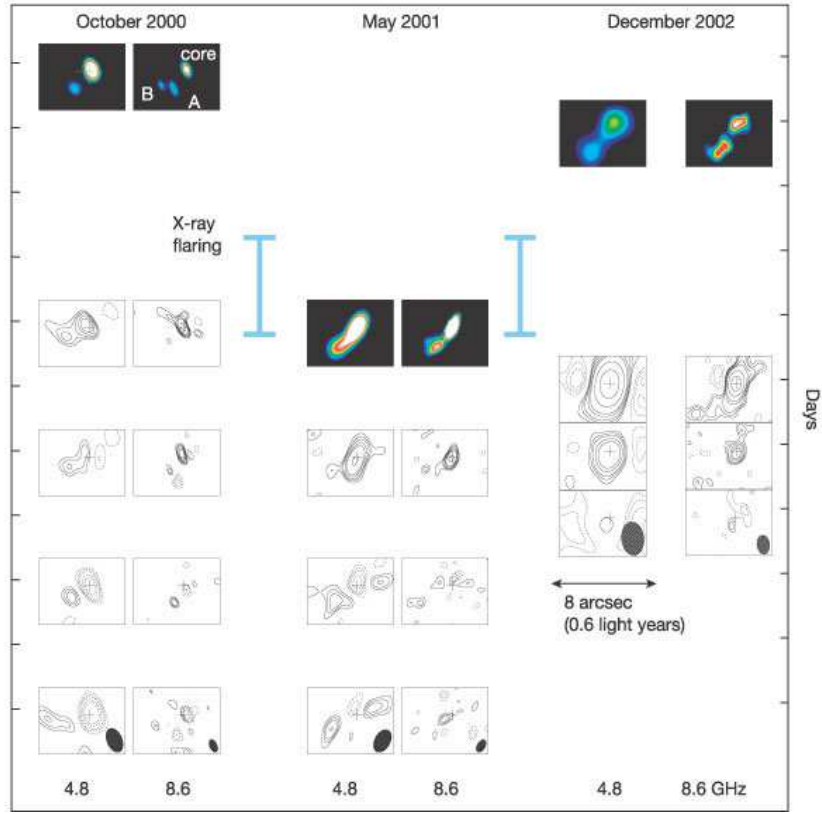} }
\caption{An ultra-relativistic jet powered by the neutron star X-ray binary
Circinus~X-1. From \cite{fendercirx1}. The inferred bulk Lorentz factor exceeds
$\Gamma=15$, making this the most relativistic Galactic jet source to date.}
\label{fig:cirx1}
\end{figure}

The Spitzer results on low-luminosity X-ray binaries (3 BHs and 1
NS;~\cite{migliari06,gallo07}) seem to point towards different energy
dissipation channels in different classes of objects, namely: the former seem
to 
be more efficient at powering synchrotron emitting outflows, having partially
self-absorbed jets which extend their spectrum up to the near-IR band.
However, it is worth reminding that the most relativistic jet discovered in the
Galaxy so far, is that the neutron star X-ray binary Circinus~X-1~\cite{fendercirx1}, for which the inferred Lorentz factor exceeds 15 (see Fig.~\ref{fig:cirx1})\index{Cir~X-1}.

\section{Jets, advection\index{advection} and event horizon\index{event horizon}}
\label{sec:horizon}

One of the most cited accomplishments of the ADAF model is to naturally
account for the relative dimness of quiescent BHBs with respect to quiescent
NS X-ray binaries~\cite{ngm97}: whereas in the case of a BH accretor all the
advected energy disappears as it crosses the horizon, adding to the BH mass,
the same energy is released and radiated away upon impact in the case of an
accretor with solid surface.  Under the ADAF working-hypothesis, such
luminosity difference -- which is indeed observed in X-rays 
\cite{garcia01} -- is actually taken as observational evidence for the
existence of an horizon in BHBs.  

However, while the observed luminosity gap may well
be a natural byproduct of advection of energy through the BH horizon, the
possibility exists that the very mechanism of jet production differs between the
two classes, and that the observed luminosity difference is due to different
channels for dissipating a common energy reservoir, with the black hole
`preferring' jets (as discussed in Sect.~\ref{sec:ns}, the NSs are indeed 
fainter than the BHs in the radio band).

\begin{figure}
\includegraphics[width=.99\textwidth]{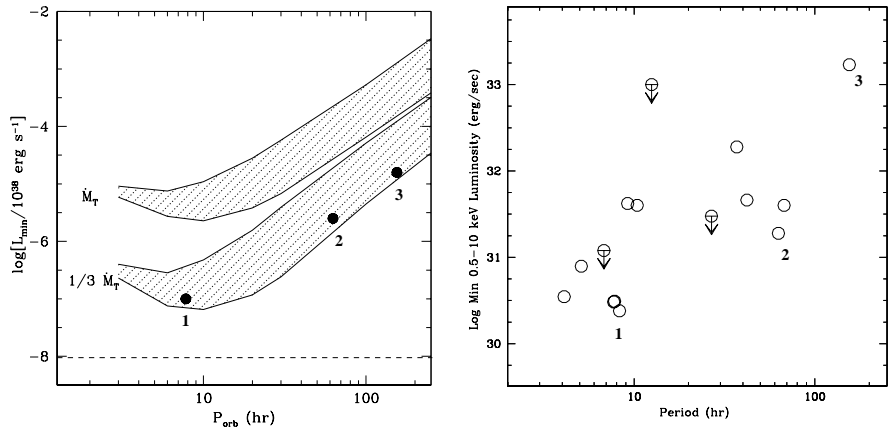} 
\caption{Left: in the ADAF scenario, the minimum luminosity around a few
$10^{30}$ \ergs~corresponds to roughly 2/3 of the outer accretion rate lost to
a wind/outflow. From \cite{menou}. Right: quiescent X-ray luminosities/upper
limits for 15 BH X-ray binaries, plotted against the systems' orbital
periods. From \cite{gallo08}.}
\label{fig:qui}
\end{figure}

Out of 15 BHB systems with sensitive X-ray observations while in the quiescent
regime, 12 have now been detected in X-rays (see Fig.~\ref{fig:qui}, right 
panel; updated 
from~\cite{tomsick03} after
\cite{corbel04,gallo06,homan06}). For those 12, the quiescent luminosities
range 
between a few $10^{30}$ and $10^{33}$ \ergs. 
The nearest BH, A~0620-00, has
been steadily emitting at $\simeq 2-3\times 10^{30}$ erg \se at least for the
past 5 years \cite{gallo06}; this is approximately the
same luminosity level as XTE~J1650-500 \cite{gallo08}, XTE~J1118+480
\cite{mcc03} and GS~2000+25 \cite{garcia01}, suggesting that this might be 
some kind of limiting value. 

In fact, for low-mass X-ray binaries one can make use of binary evolution
theory, combined with a given accretion flow solution, to predict a relation
between the minimum quiescent luminosity and the system orbital period, 
\porb~\cite{menou,lasota}.
Independently of the actual solution for the accretion flow in
quiescence, the existence of a minimum luminosity in low-mass X-ray binaries
stems directly from the existence of a bifurcation period, P$_{\rm bif}$,
below which the mass transfer rate is driven by gravitational wave radiation
($j$-driven systems), and above which it is dominated by the nuclear evolution
of the secondary star ($n$-driven systems). {As long as the luminosity expected from a
given accretion flow model scales with a positive power of the outer accretion
rate, 
systems with orbital periods close to the bifurcation period should display
the lowest quiescent luminosity}. 
As an example, the left panel of Fig.~\ref{fig:qui} (from~\cite{menou})
illustrates how the predicted luminosity of quiescent BHs powered by 
ADAFs depends on the ratio between the outer mass transfer rate and the ADAF
accretion rate. 
The lower band for instance corresponds to $\sim$1/3
of the outer mass transfer being accreted via the ADAF, {implying that the remaining 2/3
is lost to an outflow} (effectively making this inflow an ADIOS \cite{bb99}).
Interestingly, this lower band roughly reproduces the observed
luminosities of 3 representative systems spanning the whole range of detected
systems (1. A~0620-00, 2. GRO~J1655-40 and 3. V404~Cyg -- marked in both
panels).

\section{White Dwarfs}
\label{sec:wd}

\begin{figure}
\centering{\includegraphics[width=.9\textwidth]{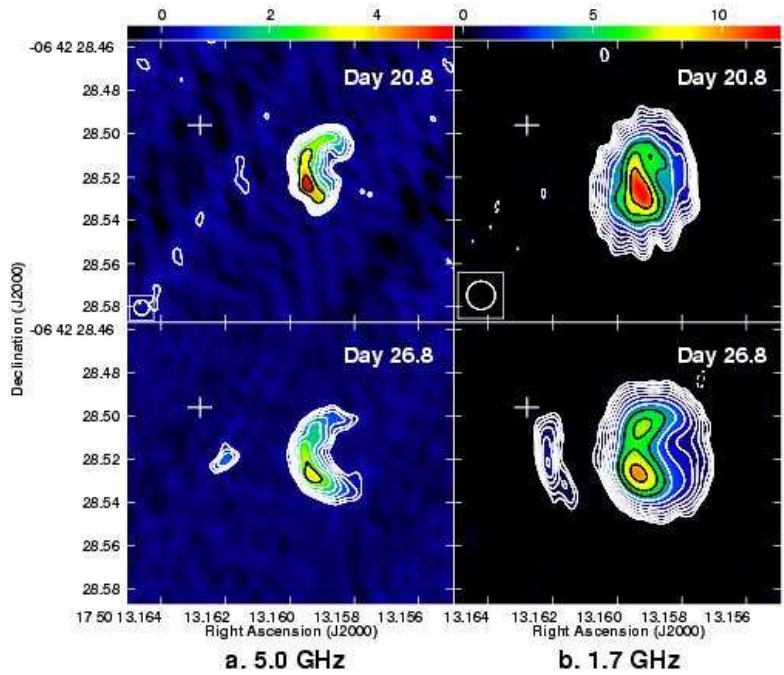}}
\caption{Two epoch, post-outburst, radio observations of the recurrent nova RS
Ophiuchi (VLBA). The second epoch observations (bottom) show evidence for
an additional resolved component to the east of the shell-like feature visible
already in the first epoch (top). From \cite{rupen07}.}
\label{fig:wd}
\end{figure}

\index{white dwarfs}
Accreting WDs come in three main classes: cataclysmic variables (CVs), supersoft X-ray
binaries, and symbiotic stars. Perhaps the best-understood of all
accretion disks are those in cataclysmic variables (CVs).
Although outflows have been observed from an increasing number of
accreting WDs in symbiotic stars and supersoft X-ray sources
\cite{Sok04,Kaw06}, these 
are generally not spatially well resolved, and questions on the jet-accretion
coupling remain open.  Most WD ejecta have velocities of hundreds to thousands
of km s$^{-1}$, and even in those cases when the emission can be spatially
separated from the binary core, it usually appears to be due to free-free
emission from shocks.  However, a handful of observations suggests that
the outflowing particles can be accelerated to high enough energies that non-thermal radio
and/or X-ray emission can also be produced \cite{Cro01,Nic07}.

\begin{figure}[t!]
\centering{\includegraphics[width=.99\textwidth]{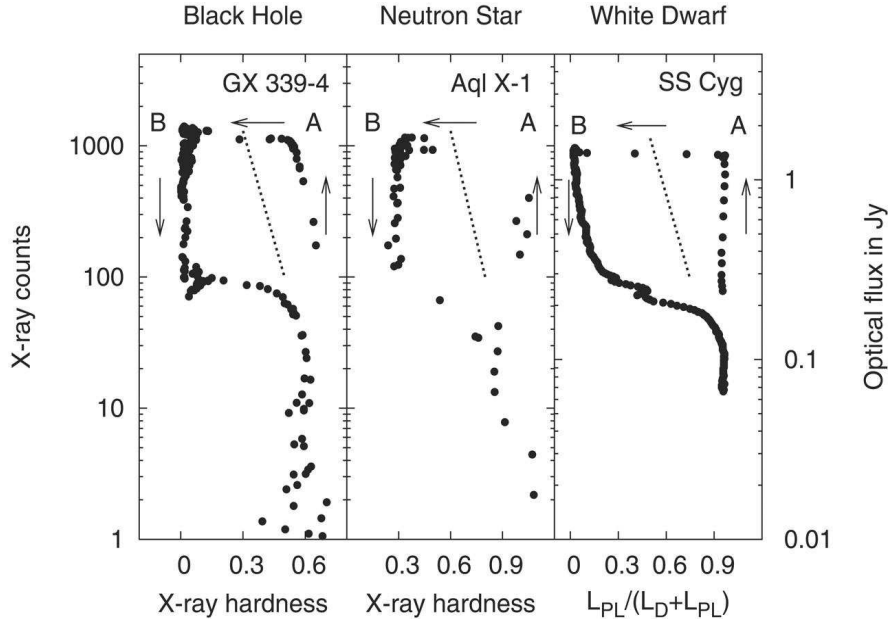} }
\caption{Disk-fraction vs. luminosity diagrams for Galactic binary stellar
systems and microquasar sources hosting different compact objects: a BH, NS
and WD. From \cite{elmarscience}. This study provides the strongest observational
evidence to date for a synchrotron-emitting outflow from a dwarf nova system,
and strengthens the similarities between jet sources across the whole mass
spectrum.}
\label{fig:wdhid}
\end{figure}

Two recent works~\cite{rupen07,elmarscience} have provided us with the best observational evidence for
synchrotron emission from a WD jet. 
As illustrated in Fig.~\ref{fig:wd}, VLBA observations
of the recurrent nova RS~Ophiuchi\index{RS~Oph}, taken about 21 and 27 days after the
outburst peak, have imaged a resolved synchrotron component well to the
east of the shell-like shock feature\footnote{Typically, post-outburst radio
emission from RS~Ophiuchi originates from dense ISM that is swept up,
compressed and shocked by the relatively rarefied 
shell that has been ejected ~\cite{sea89}}. Interestingly, the inferred jet
velocity is comparable to the escape velocity from a WD.

A second example is the detection of a transient radio jet erupting from the
dwarf nova SS~Cyg \cite{elmarscience}\index{SS~Cyg}. Radio observations of this system,
conducted during its 2007 April outburst with the VLA, revealed a variable,
flat-spectrum source with high brightness temperature, most likely due to
partially self-absorbed synchrotron emission from a jet \cite{elmarscience}.
As apparent from Fig.~\ref{fig:wdhid}, the behavior of SS~Cyg (right panel)
during the outburst closely resembles that of X-ray binaries hosting
relativistic compact objects. Plotted here, from left to right, are the
`disk-fraction vs. luminosity diagrams' (which can be seen as the
generalization of HIDs generally adopted for X-ray binaries; see \cite{kjf})
for three different accreting objects: a BH, NS and the WD SS~Cyg. Unlike for
BHs and NSs, where the x-axis is the X-ray spectral hardness,
for the WD case this variable has been replaced by the power-law fraction,
which quantifies the prominence of the power law component over the boundary
layer/thermal disk component of the energy spectrum, as measured in the
optical/UV band.  Evidently, the three systems occupy the same region of the
parameter space during a typical outburst (see Fig.~\ref{fig:turtle} for a
schematic of typical BH outburst HID, and Fig.~\ref{fig:ns}, bottom left panel, for
Z-type NSs), the most notable difference being that -- for the BHs -- the
core, flat spectrum radio emission drops below detectable levels when in the
soft/thermal dominant state, while it seems to be only somewhat weakened in
NSs and WDs.

These results provide further support to the hypothesis that the mere existence of accretion coupled to
magnetic fields may be sufficient ingredients for a jet to form, independently
of the presence/absence of an horizon.
However, the latter (or, put it another way, the presence of a physical
surface) could play a role in: {\it i)} making the jets more radio loud;  {\it
ii)} enhancing the magnitude of the jet suppression mechanism during bright
soft states. 

\section{Jet production, collimation\index{collimation}, matter content}
\label{sec:unknown}
It is often
assumed that the velocity the the steady jet is only mildly
relativistic, with $\Gamma \simeq 2$~\cite{gfp}. This comes from the
relative spread about the radio/X-ray correlation, interpreted as
evidence for a low average Lorentz factor. However, this argument has been confuted on theoretical
grounds~\cite{hm04}.  As far as the transient jets are concerned,
there is a high degree of uncertainties in estimating their Lorentz
factors, mainly because of distance uncertainties~\cite{fender03}.

\begin{figure}[t!]
\includegraphics[width=.55\textwidth]{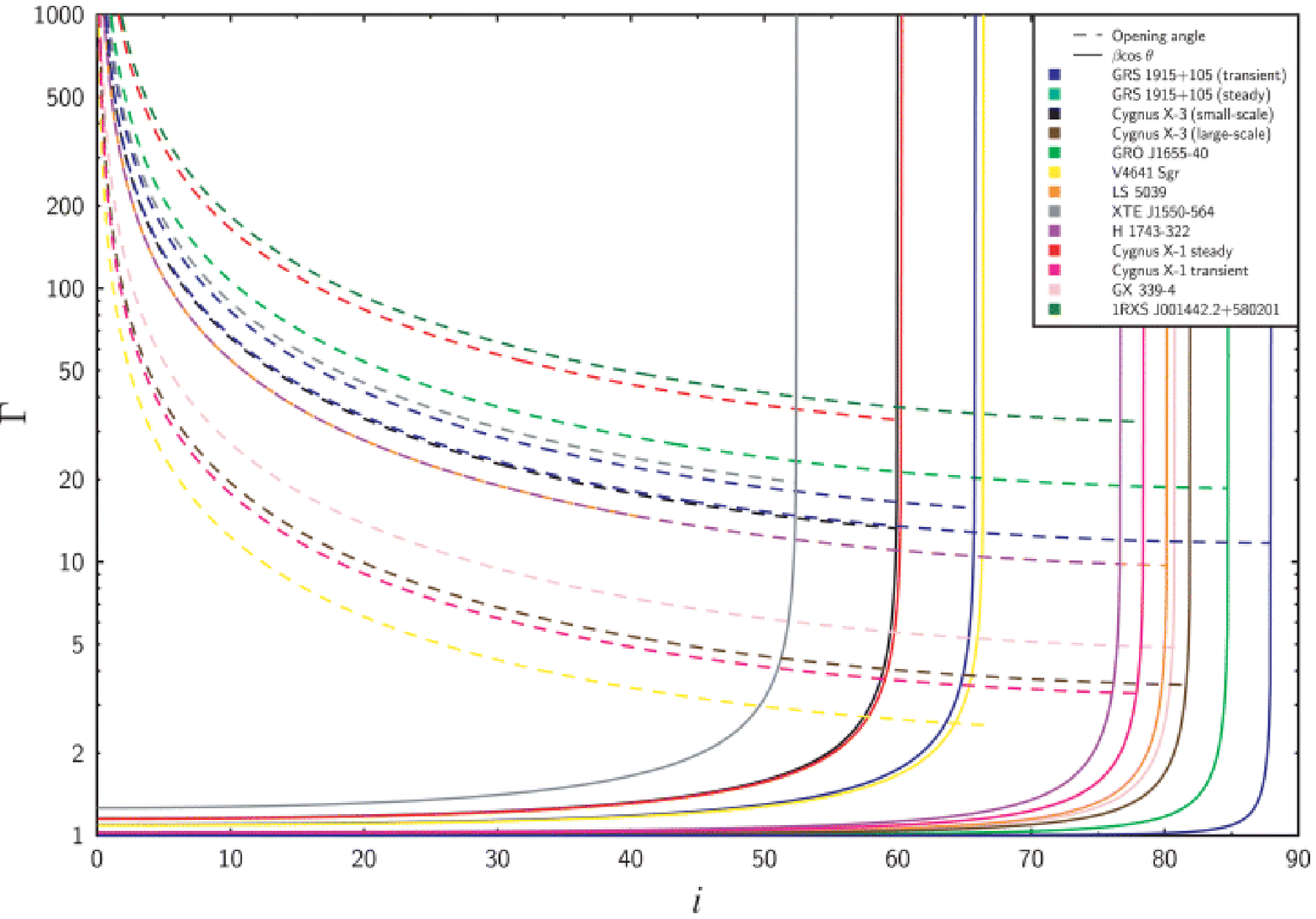}\includegraphics[width=.4\textwidth]{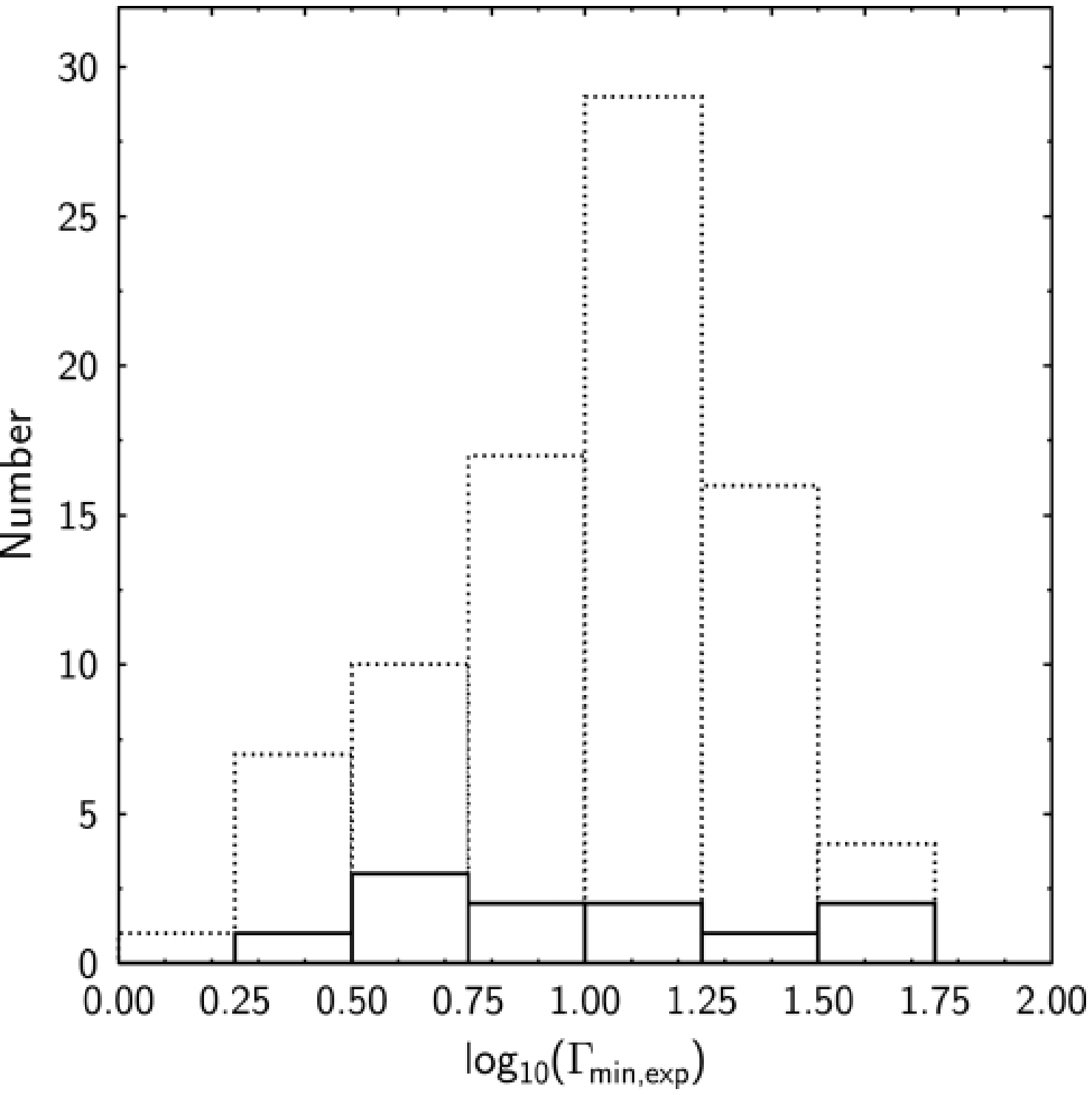}
\caption{Left:  Lorentz factors derived from opening angles (Eq. 4 in~\cite{jmj06}, dashed lines) and from the inferred $\beta\times {\rm cos}~i$ products 
for a number of microquasars. Since the Lorentz factors are derived from upper limits on the opening angles they are in fact lower limits (assuming freely-expanding jets with an expansion speed $c$). Right: Mean Lorentz factors for X-ray binaries from opening angles'
constraints (solid), compared to AGN from proper motions'
(dashed). From~\cite{jmj06}.}
\label{fig:jmj}
\end{figure}

A recent work~\cite{jmj06} made a substantial step forward in constraining the
Lorentz factor of microquasars by means of the \index{Lorentz factor}
observational upper limits on the jets' opening angles.  This method relies on
the fact that, while the jets could undergo transverse expansion at a
significant fraction of the speed of light, time dilation effects associated
with the bulk motion will reduce their apparent opening angles.  \cite{jmj06}
have calculated the Lorentz factors required to reproduce the small opening
angles that are observed in most X-ray binaries, with very few exceptions,
{under the crucial assumption of no confinement}.  The derived values, mostly
lower limits, are larger than typically assumed, with a mean $\Gamma_{\rm m}
>$10. No systematic difference appears to emerge between hard state steady
jets and transient plasmons (Fig. \ref{fig:jmj}).  If indeed the transient
jets were as relativistic as the steady jets, as already mentioned, this would
challenge the hypothesis of internal shocks at work during hard-to-thermal
state transitions in BHBs. In order for that scenario to be viable, the
transient jets must have higher Lorentz factor; in other words, steady jets
ought to be laterally confined.

\begin{figure}[b!]
\centering{\includegraphics[width=.8\textwidth]{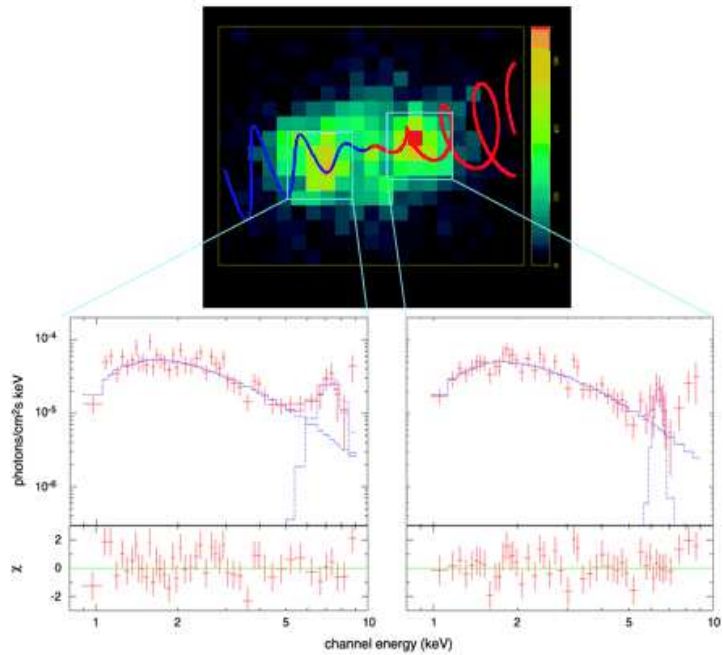}}
\caption{Chandra image of SS~433 with the projected precession cycle of the
jets superimposed on it. These observations reveal evidence for a hot
continuum and Doppler-shifted iron emission lines from spatially resolved
regions. From \cite{mfm02}.} 
\label{fig:ss433}
\end{figure}

The issue of the jets' matter content remains highly debated. Perhaps with the
exception of SS~433 \cite{watson,mfm02}, where atomic lines have been
detected at optical at X-ray wavelengths along the jets (see
Fig.~\ref{fig:ss433}), various studies come to different conclusions for
different sources. For example, circular polarization, which in principle can
provide an excellent tool for investigating the baryonic content of the jets,
is only detected in a handful of sources~\cite{fenderpol}, where no strong
conclusion could be placed yet.  Entirely different studies (e.g. based on
modeling large scale jet-ISM interaction structures by means of self-similar
jet fluid models) also draw different conclusions: in the case of Cygnus~X-1
for instance, some authors \cite{gallo05,heinz06} argue for cold baryons in
the flow, while an electron/positron jet seems to be favored in GRS 1915+105
based on energetics arguments~\cite{fenderpol}.

\section{Future prospects}
\label{sec:future}

In this last Section, I choose to briefly introduce a number of observational 
capabilities that are about to become available to the community -- or have
just started to -- that I believe will
literally revolutionize our understanding of microquasars (and not only).

\subsection{Next generation radio interferometers}

With an imaging resolution of 50 milliarcsec at 40 GHz, the imaging capability
of the VLA is comparable with the highest resolution of Advanced Camera for
Surveys on-board the Hubble Space
Telescope. However, the fundamental data-processing capabilities of the VLA
remained essentially unchanged since the 70s'. The Expanded VLA
(EVLA\footnote{\url{http://www.aoc.nrao.edu/evla/}}) will dramatically improve
its ability to make high-sensitivity and high-resolution images.  The {EVLA} \index{EVLA}
will attain unprecedented image quality
with 10 times the sensitivity and 1000 times the spectroscopic capability of
the existing array. Finally, the addition of eight new antennas will provide an
order-of-magnitude increase in angular resolution. 
At the time of writing, the number of EVLA antennas continues to increase at a rate of one every two months. 
Each upgraded EVLA antenna produces 100 times more data than an original VLA antenna. When completed in 2012, the EVLA will be the most powerful centimeter-wavelength radio telescope in the world. The technology developed for the EVLA will enable progress on the next generation radio telescope: the Square Kilometer Array.

With baselines of up to 217 km, the Multi-Element Radio Linked Interferometer
Network (MERLIN) provides cm-wavelength imaging at 10 to 150 milliarcsec
resolution, effectively covering the gap between arrays such as the Westerbork
Synthesis Radio Telescope (WSRT) and the VLA, and Very Long Based
Interferometry (VLBI) arrays such as the VLBA and the European VLBI Network
(EVN). e-MERLIN\footnote{\url{http://www.jb.man.ac.uk/research/rflabs/eMERLIN.html}}
is a major UK project aimed at increasing the bandwidth and thus the sensitivity of
MERLIN by about an order of magnitude. This increased sensitivity,
together with the high resolution provided by the long baselines,
will enable a wide range of new astronomical observations, including of course
Galactic microquasars. 
One other major
development which is part of the upgrade is frequency flexibility, as 
e-MERLIN will be able to switch rapidly
between 1.4, 5, 6 and 22 GHz.  The e-MERLIN\index{e-MERLIN} 
telescope array is now nearing completion, and will soon start to acquire data for its approved legacy programs.\\

In terms of microquasar studies, the EVLA and e-MERLIN will be simply a
revolution. They will allow for the detection of faint quiescent systems in
short exposures, enabling us to test whether collimated jets survive in
quiescent black holes (such as V404~Cyg). It will be possible to perform
systematic searches for radio counterparts to NSs. Radio outbursts will
hopefully receive the same daily coverage as X-rays outbursts do, and with
amazingly fast frequency switching capabilities. Deep searched for low surface
brightness jet-powered nebulae will be carried out at relatively
(comparatively) limited expenses in
terms of telescope time. Finally, we will be able to perform spectroscopic
studies of microquasars jet, search for lines, possibly constrain the jet
baryon content {\emph{ directly}}.

{LOFAR}\footnote{\url{http://www.lofar.org/}} (Low Frequency Array) is a\index{LOFAR}
next-generation radio telescope currently under construction in The Netherlands, with
baseline stations under development over a number of EU countries. The array will operate in the
30-80 and 120-240 MHz bands, thus representing the largest of the pathfinders
for the lowest-frequency component of the Square Kilometer Array (SKA)
project (see \cite{fenderlofar}). Core
Station One of LOFAR is currently operating; according to the plan, 36
stations will be deployed by the end of 2009.  

LOFAR will literally 
revolutionize the study of bursting and transient radio phenomena at low radio
frequencies. 
So far, the primary instruments for detecting extragalactic Gamma-ray bursts and galactic
microquasars have been orbiting satellites, such as BeppoSAX, INTEGRAL, RXTE, 
{Swift} and others. One of the breakthroughs in this field was the 
localization of Gamma-ray burst and their subsequent identification with
supernovae/galaxies at high red-shifts. From the empirical relation between radio
and X-ray emission for these systems (see Chap. 5) it is apparent that the all-sky monitoring
with LOFAR will be a factor of 5-10 more effective in discovering such events
than previous all-sky-monitors \cite{fenderlofar}. Furthermore, LOFAR will allow much more accurate
localization of these events, enabling fast response science and follow-ups at
other wavelengths.

\subsection{$\gamma$-ray binaries}

A new observational window has just been opened by ground-based Cherenkov
telescopes, HESS\footnote{\url{http://www.mpi-hd.mpg.de/hfm/HESS/HESS.html}}
(High Energy Stereoscopic System) and
MAGIC\footnote{\url{http://wwwmagic.mppmu.mpg.de/}} (Major Atmospheric
Gamma-ray Imaging Telescope), that survey the sky above 100 GeV. Because of
their high sensitivity, and high angular and energy resolution, these
telescopes are revealing and identifying a plethora of new extragalactic and
galactic sources of very-high energy ($>$100 GeV) radiation. The Galactic
Center, supernovae remnants, pulsar-wind nebulae, and some `$\gamma$-ray
binaries' have all been identified as very high $\gamma$-ray sources in the
Galaxy~\cite{mirabel06}.  The HESS collaboration reported the detection of TeV
$\gamma$-ray emission from the Be type binary system PSR~B1259-63\index{PSR~B1259-63}, close to
the periastron passage~\cite{aharonian05a}. Also TeV $\gamma$-rays have been
detected from LS~5039~\cite{aharonian05b} and LS~I~+61~303~\cite{albert06}. 
\index{LS~5039}\index{LS~I~+61 303}
The idea of Galactic jet sources as capable of 
accelerating particles up to very high energies has been strengthened by the
direct detection of large scale X-ray jets resulting from in-situ shocks where
electrons are accelerated up to TeV energies
\cite{corbel02,corbel05}. 
The synergy between very high energies and lower
frequency observations is exemplified by the recent detection of a
$\gamma$-ray flare from the BH X-ray binary Cygnus~X-1, reported by the MAGIC
collaboration. The flare, seen simultaneously with
RXTE, Swift, and INTEGRAL, was compatible with a point-like source at
a position consistent with the binary system, thus ruling out its
arcmin scale jet-driven radio nebula~\cite{albert07}.
Alternatively, relativistic particles can be injected in the
surrounding medium by the wind from a young pulsar (see e.g. the case of 
LS~I~+61~303~\cite{dhawan06}).  { Coordinated
multi-wavelength monitoring can discriminate between competing models,
as they predict different emission regions for the radio and
$\gamma$-ray radiation (core vs. arc-minute scale $\gamma$ emission
can now be resolved by high resolution $\gamma$-ray facilities), as
well as different flux variations with the orbital phase}.

Finally, scheduled to launch in mid 2008, the Gamma-ray Large Area Space
Telescope\footnote{\url{http://glast.gsfc.nasa.gov/}} (GLAST) is an
international and multi-agency space mission which will study the cosmos in
the energy range 10 keV - 300 GeV, complementing ground-based Cherenkov
telescopes with wider field.  In the 90s, EGRET (the Energetic Gamma Ray
Experiment Telescope) made the first complete survey of the sky in the 30
MeV--10 GeV range, showing the $\gamma$-ray sky to be surprisingly dynamic and
diverse. {\em Most} of the EGRET sources remain unidentified (170 over 271):
this outlines the importance and potentials for new discoveries of the GLAST
mission, whose Large Area Telescope (LAT) has a field of view about twice as
wide (more than 2.5 steradians), and sensitivity about 50 times at 100 MeV.

The future is bright.\\
 
\noindent
{\bf Acknowledgments:} The author is supported by NASA through a
Hubble Fellowship grant HST-HF-01218 issued from the Space Telescope Science Institute, which is
operated by the Association of Universities for Research in Astronomy,
Incorporated, under NASA contract NAS5-26555.
I wish to thank Rob Fender,
Elmar K{\"o}rding, James Miller-Jones and Valeriu Tudose for providing me
with original figures, in part still unpublished.

\end{document}